\documentclass[journal]{IEEEtran}

\usepackage[pdftex]{graphicx}
\usepackage[cmex10]{amsmath}
\usepackage[tight,footnotesize]{subfigure}
\usepackage[font=footnotesize]{subfig}
\usepackage{url}
\usepackage{url}

\begin{document}
\title{Impact of Gate Assignment on Gate-Holding Departure Control Strategies}

\author{Sang~Hyun~Kim,
        and~Eric~Feron
\thanks{S. H. Kim is with the School of Aerospace Engineering, Georgia Institute of Technology, Atlanta, GA 30332 USA (e-mail: sanghyun.kim@gatech.edu)}
\thanks{E. Feron is with the School of Aerospace Engineering, Georgia Institute of Technology, Atlanta, GA 30332 USA, and with Ecole Nationale de l'Aviation Civile, 31000 Toulouse, France (e-mail: feron@gatech.edu).}
}

\maketitle

\begin{abstract}
Gate holding reduces congestion by reducing the number of aircraft present on the airport surface at any time, while not starving the runway. Because some departing flights are held at gates, there is a possibility that arriving flights cannot access the gates and have to wait until the gates are cleared. This is called a gate conflict. Robust gate assignment is an assignment that minimizes gate conflicts by assigning gates to aircraft to maximize the time gap between two consecutive flights at the same gate; it makes gate assignment robust, but passengers may walk longer to transfer flights. In order to simulate the airport departure process, a queuing model is introduced. The model is calibrated and validated with actual data from New York La Guardia Airport (LGA) and a U.S. hub airport. Then, the model simulates the airport departure process with the current gate assignment and a robust gate assignment to assess the impact of gate assignment on gate-holding departure control. The results show that the robust gate assignment reduces the number of gate conflicts caused by gate holding compared to the current gate assignment. Therefore, robust gate assignment can be combined with gate-holding departure control to improve operations at congested airports with limited gate resources.
\end{abstract}


%
\IEEEpeerreviewmaketitle

%
%

\section{Introduction}
\IEEEPARstart{G}{ate} holding is an approach to reduce taxi delays and emissions in the departure process while maintaining airport departure throughput (take-off rate), which is motivated by the fact that the number of take-offs per minute is saturated when the number of aircraft that taxi out (denoted $N$) is greater than a saturation point ($N^*$) \cite{shumsky1995, anagnostakis2000conceptual}. Shumsky, in his dissertation \cite{shumsky1995}, notices that the take-off rate rises with $N$ and levels off a certain value after $N$ is over $N^*$. As a result, the take-off rate corresponding to $N^*$ and above represents the airport's departure capacity. Gate holding manages to keep $N$ near $N^*$ by controlling pushback clearances. When $N$ exceeds $N^*$, gate holding becomes active and aircraft requesting pushback clearance are held at gates. According to the recent study \cite{jung2011performance}, taxi delays in the departure process are transformed to gate holding delays by utilizing gate-holding departure control, without sacrificing airport capacity. Gate holding was implemented experimentally at Boston Logan Airport following the development of suitable human interfaces \cite{simaiakis2011demonstration}, and it was shown that gate holding helps the airport system to shift taxi-out times to environmentally and financially less expensive gate-holding times. One issue is whether this gate-holding strategy can be detrimental to the free access of arriving flights to the terminals. This is particularly true for congested and resource-limited airports such as La Guardia Airport (LGA) and Hartsfield-Jackson Atlanta International Airport (ATL). Recent studies relating to gate holding by Jung et al. \cite{jung2011performance} and Simaiakis et al. \cite{simaiakis2011demonstration} indicate that the research community is still concerned with consequences of gate holding on terminal airside congestion.

This paper investigates the impact of smart gate assignment on the gate holding strategies described above. Gate holding addresses airport surface congestion by holding an aircraft at its gate, thus taking advantage of the time gap between consecutive gate uses. This gap, which we call gate separation, can constrain the efficiency of gate holding. For instance, when an aircraft is held at the gate and an arriving aircraft requests the same gate, either the gate-held aircraft must be cleared for pushback, or the arriving aircraft must wait for the gate hold to terminate. In both situations, gate holding is prevented from working fully. This study sheds light on the importance of understanding the impact of gate assignment on gate holding and vice-versa.

Section II presents the airport departure model, a queuing model that consists of a take-off model and taxi-out time estimators. The model is calibrated and validated with historical data. Section III presents the airport gate assignment problem for robust gate assignments. Section IV analyzes the impact of gate assignment on gate-holding departure control, and Section V concludes and summarizes the findings.

\section{Airport Departure Model}
\subsection{Queuing Model}
Many researchers use a queuing model for simulating the airport departure process \cite{pujet2003input, simaiakis2010queuing, burgain2011valuating}. The queuing models have a similar structure: When an aircraft is ready for push-back, it enters a push-back queue. When gate holding is inactive, the push-back is cleared on a First-Come-First-Served (FCFS) basis. However, if gate holding is active, a push-back is cleared when the number of aircraft on the ramp or taxiway system is below a critical number $N^*$. After the aircraft is cleared for push-back, the taxi-out time to a runway threshold is generated. When the aircraft reaches the runway threshold, it enters a runway queue and is cleared for take-off on a FCFS basis. 

There are some research efforts to simulate more detailed aircraft motion on the airport surface. For instance, NASA developed a high fidelity human-in-the-loop simulation model, and the simulation model is used to assess their Spot and Runway Departure Advisor tool \cite{hoang2011tower, jung2011performance}. Such a detailed simulation-based model can capture congestions and queues at all potential queuing locations such as taxiway merge-locations. The queuing model used in this paper and discussed above has a limited capability to simulate potential queues on the airport surface. Indeed, the queuing model has queues only in the ramp area and the runway. However, the queuing model used in this paper enables faster simulations than detailed models, and it is capable of accurately estimating push-back and take-off time distributions. The objective of this paper is to analyze the impact of gate assignment on gate-holding departure control, and the resolution of the queuing model is sufficient for the analysis. Therefore, the queuing model is used for the simulation in this paper. 

\subsection{Data Source}
\begin{figure}[!t]
	\centering
	\includegraphics[width=2.5in]{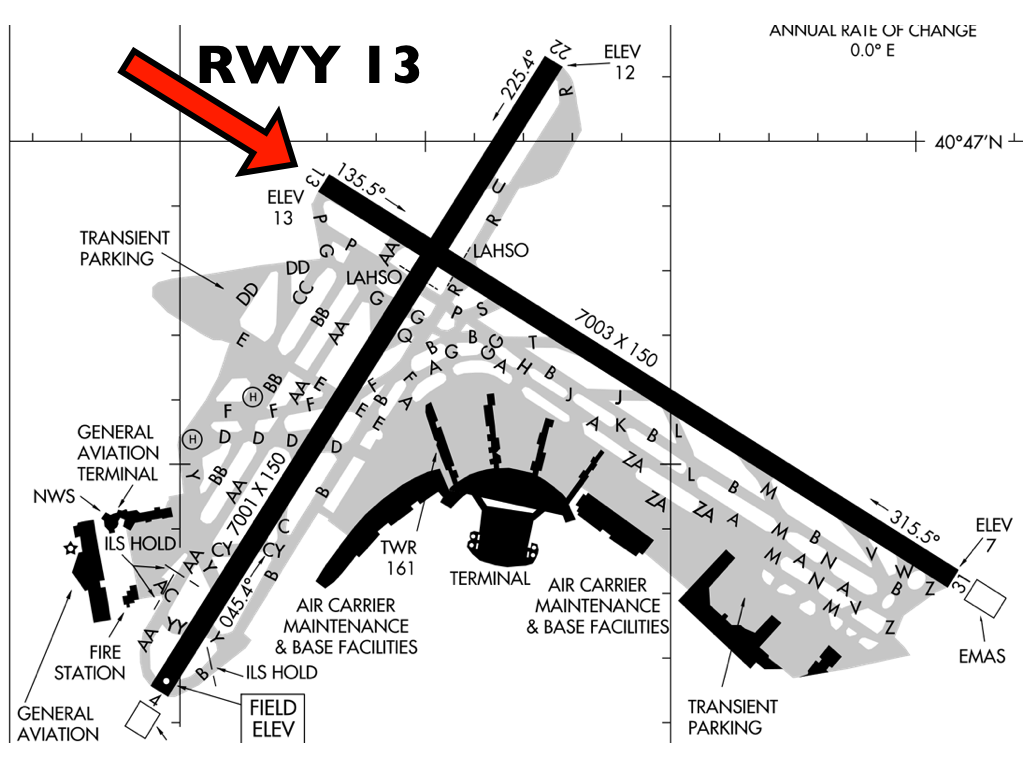}
	\caption{Layout of La Guardia Airport and Runway 13}
	\label{f:lga}
\end{figure}

The queuing model is calibrated to LGA operations using 2009 data from Aviation System Performance Metrics (ASPM) provided by the Federal Aviation Administration (FAA). ASPM contains actual departure time, take-off time, taxi-out time, tail number, runway configuration, etc. The data is categorized by departure runway. Frequently used runway configurations are given in Table~\ref{t:conf}, and the layout of LGA is shown in Fig.~\ref{f:lga}. Most of the time, one runway is used for arrivals and another runway is used for departures. Table~\ref{t:conf} shows that runway 13 served departures the most frequently. Precisely, runway 13 operated for 3456 hours (39.5\% of the year) and served 83143 push-backs (47.6\% of push-backs that year). So, the queuing model is calibrated with departures from runway 13.

\begin{table}[!t]
	\caption{Frequently Used Runway Configuration in LGA}
	\label{t:conf}
	\centering
	\begin{tabular}{|c|c|c|}
		\hline
		Configuration (Arrival | Departure) & \% of Hours & \% of Pushbacks \\ \hline \hline
		31 | 4 & 17.4 \% & 22.6 \% \\ \hline
		4 | 13 & 15.4 \% & 19.4 \% \\ \hline
		22 | 31 & 14.0 \% & 18.2 \% \\ \hline
		13, 22 | 13 & 12.0 \% & 14.9 \% \\ \hline
		22 | 13 & 7.7 \% & 10.1 \% \\ \hline
	\end{tabular}
\end{table}

\subsection{Take-off Model}
\begin{figure}[!t]
	\centering
	\includegraphics[width=2.5in]{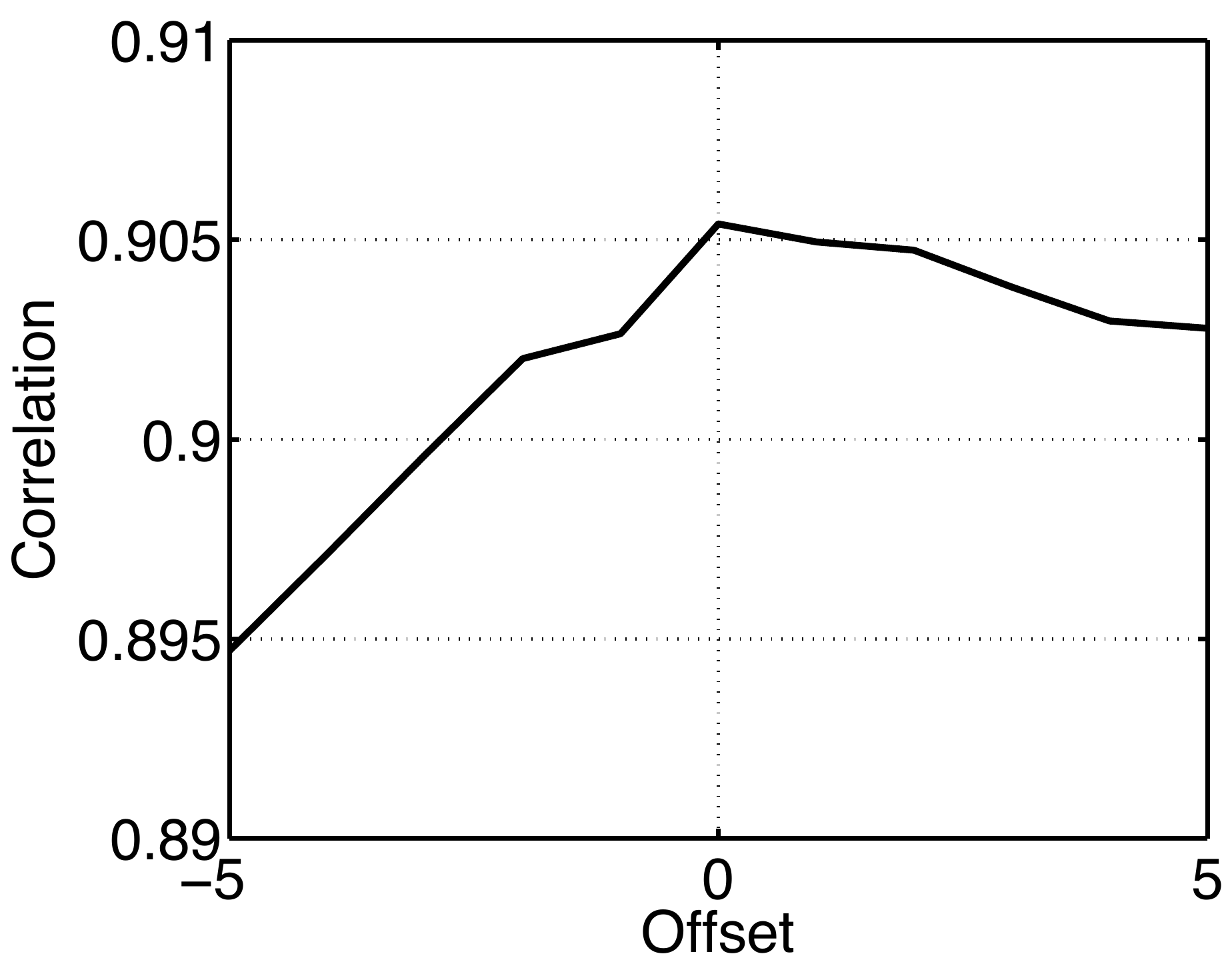}
	\caption{Correlation between $N(t)$ and $T(t+\delta t)$}
	\label{f:NTcorr}
\end{figure}

As Shumsky has shown in \cite{shumsky1995}, the take-off rate is related to $N$, the number of aircraft on the airport surface. Let $N(t)$ be the number of taxi-out aircraft at time $t$ and $T(t)$ be the average number of take-offs per minute over the time periods $[t, t+9]$. Pujet et al. calculate the correlation between $N(t)$ and $T(t+\delta t)$ in order to find $\delta t$, where $N$ predicts accurate $T$ \cite{pujet2003input}. Fig.~\ref{f:NTcorr} shows that the maximum correlation between $N$ and $T$ occurs with $\delta t = 0$. Therefore, the number of taxi-out aircraft at an instant of time predicts the number of take-offs over the next 10 minutes. 

\begin{figure}[!t]
	\centering
	\includegraphics[width=2.5in]{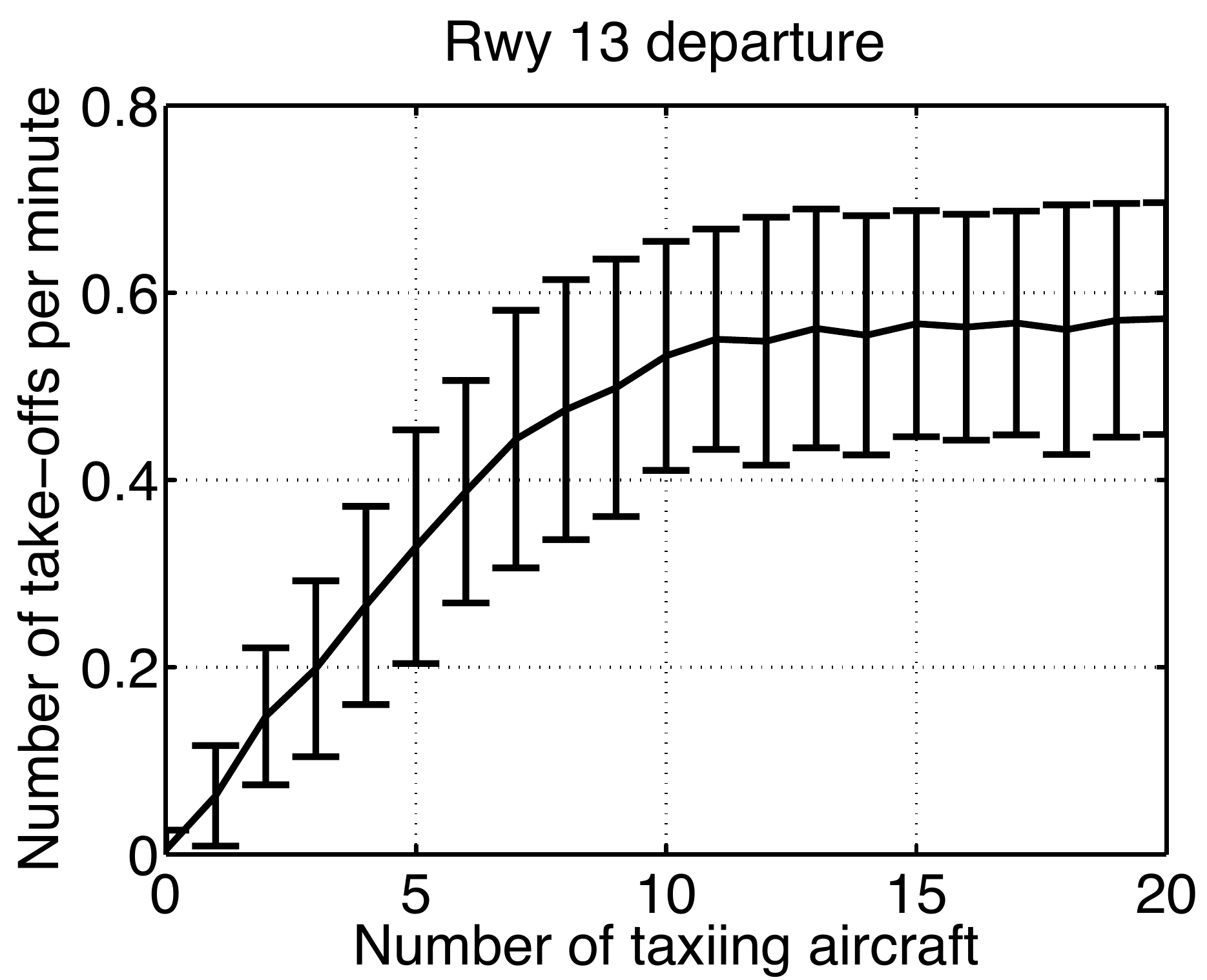}
	\caption{$T(t)$ as a Function of $N(t)$: The Vertical Bars Indicate the Standard Deviation of $T(t)$ for Each $N(t)$.}
	\label{f:NT}
\end{figure}

Fig.~\ref{f:NT} shows average $T(t)$ according to $N(t)$, and the vertical bars indicate the standard deviation of $T(t)$ for each $N(t)$. $T(t)$ and $N(t)$ are calculated by analyzing ASPM data. $T(t)$ increases with $N(t)$ until $N(t)$ becomes 15, which is $N^*$. When $N$ is greater than or equal to $N^*$, the departure throughput is limited by the runway capacity. The runway capacity is obtained from $T(t)$ when $N(t)$ is in the range of [15, 20]. The mean and standard deviation of take-off rate is 0.5666 aircraft per minute and 0.1234 aircraft per minute. Fig.~\ref{f:NT} is similar to the flow-density curve in the ground transportation literature \cite{daganzo1994cell, wong2002multi}. Unlike road models where throughput (flow) is limited by road characteristics such as jam density and free flow speed, airport throughput is limited by runway capacity only. In some extreme cases, airport throughput can be seen to decrease, like in the road models, for high number of taxiing aircraft (airport aircraft density). Such gridlock situations are, however, rare. 

In order to simulate the first and the second moment of $T(t)$, two variables ($p_1$ and $p_2$) and three parameters ($c_1$, $c_2$, and $c_3$) are defined. A take-off clearance is modeled as follows:
\begin{itemize}
	\item $c_1$ aircraft per minute with probability $p_1$,
	\item $c_2$ aircraft per minute with probability $p_2$,
	\item $c_3$ aircraft per minute with probability $1-p_1-p_2$.
\end{itemize}
Then, the runway capacity is expressed
\begin{equation}
	\label{e:mu}
	\mu = c_1 p_1 + c_2 p_2 + c_3 (1-p_1-p_2)
\end{equation}
\begin{equation}
	\label{e:sig}
	\sigma = \sqrt{\frac{c_1^2 p_1 + c_2^2 p_2 + c_3^2 (1-p_1-p_2) - \mu^2}{10}}.
\end{equation}
The three parameters are explored in increments of 0.025 to find the best set of parameters that simulates the distribution of take-off rate at $N=N^*$. The variables $p_1$ and $p_2$ are calculated by (\ref{e:mu}) and (\ref{e:sig}). The variables and parameters of the take-off model are given in Table~\ref{t:param}.

\begin{table}[!t]
	\caption{Variables and Parameters of the Take-off Model}
	\label{t:param}
	\centering
	\begin{tabular}{|c|c|}
		\hline
		Name & Value \\ \hline \hline
		$c_1$ & 0.525 aircraft/minute \\ \hline
		$c_2$ & 1.025 aircraft/minute \\ \hline
		$c_3$ & 0.025 aircraft/minute \\ \hline
		$p_1$ & 0.3733 \\ \hline
		$p_2$ & 0.38 \\ \hline
	\end{tabular}
\end{table}

\begin{figure}[!t]
	\centering
	\includegraphics[width=2.5in]{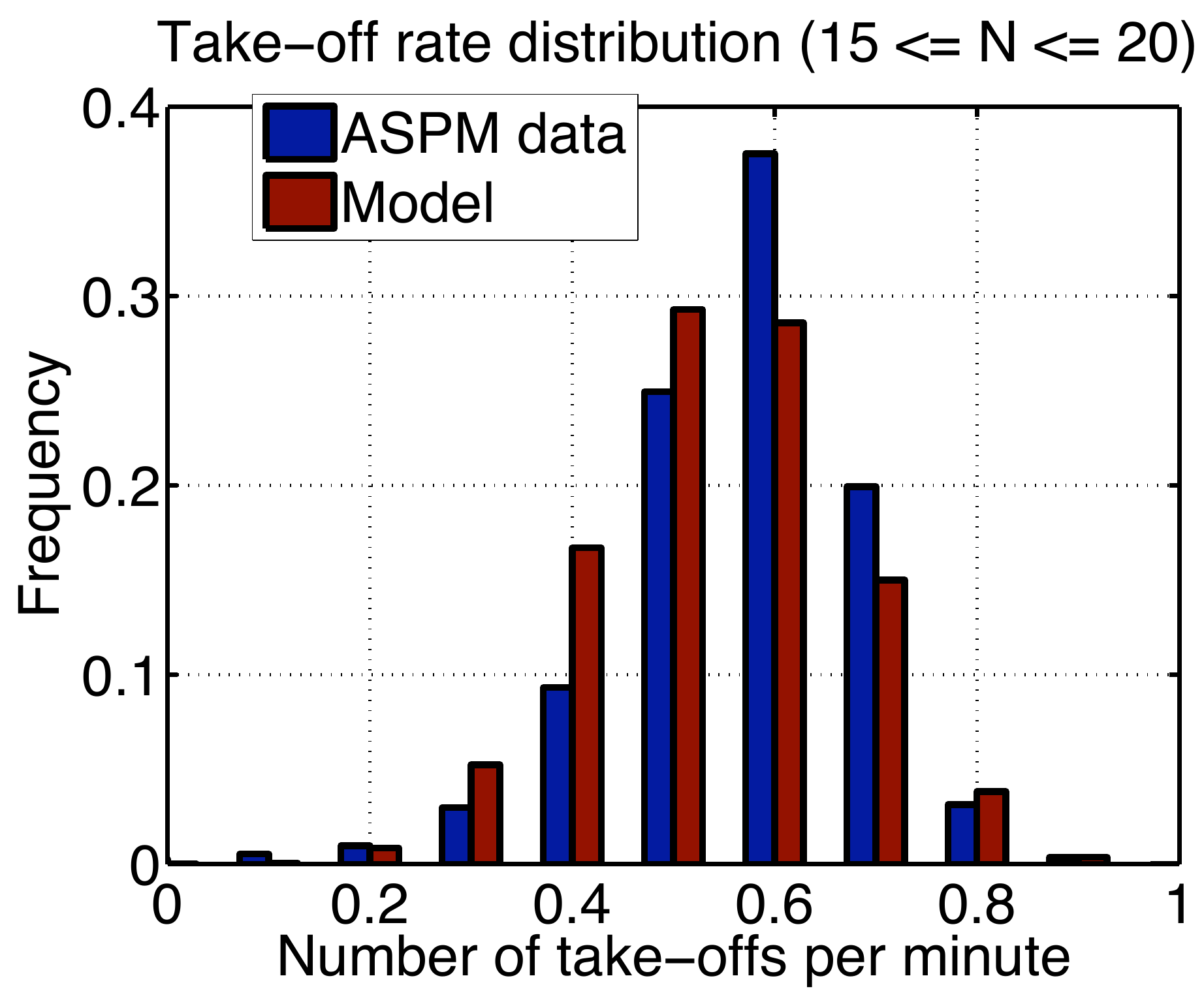}
	\caption{Take-off Rate Distribution}
	\label{f:takeoff}
\end{figure}

When the runway queue is not empty, the take-off model randomly selects the current take-off rate $c$ according to $p_1$ and $p_2$. So, $c$ is equal to $c_1$, $c_2$, or $c_3$ with probabilities $p_1$, $p_2$, or $1-p_1-p_2$, respectively. The current take-off rate $c$ is added to the previous take-off rate, and the largest integer smaller than this cumulative sum of $c$ is the maximum number of take-off clearances at the current time step. Then, the actual number of take-offs is subtracted from the cumulative sum of $c$. For example, the cumulative sum of $c$ at the previous time step is 0.55, and the $c$ at the current time step is 0.525. Then, the cumulative sum of $c$ at the current time step becomes $0.55+0.525=1.075$, and one take-off can be cleared at the current time step. Because there are aircraft in the runway queue, the take-off model clears a take-off. Then, the cumulative sum of $c$ becomes $1.075-1=0.075$. Fig.~\ref{f:takeoff} shows the distribution of take-off rates from ASPM data and the take-off model. The throughput of the airport departure model is sensitive to the take-off model.

\subsection{Taxi-out Time Estimator}
\begin{figure}[!t]
	\centering
	\includegraphics[width=2.5in]{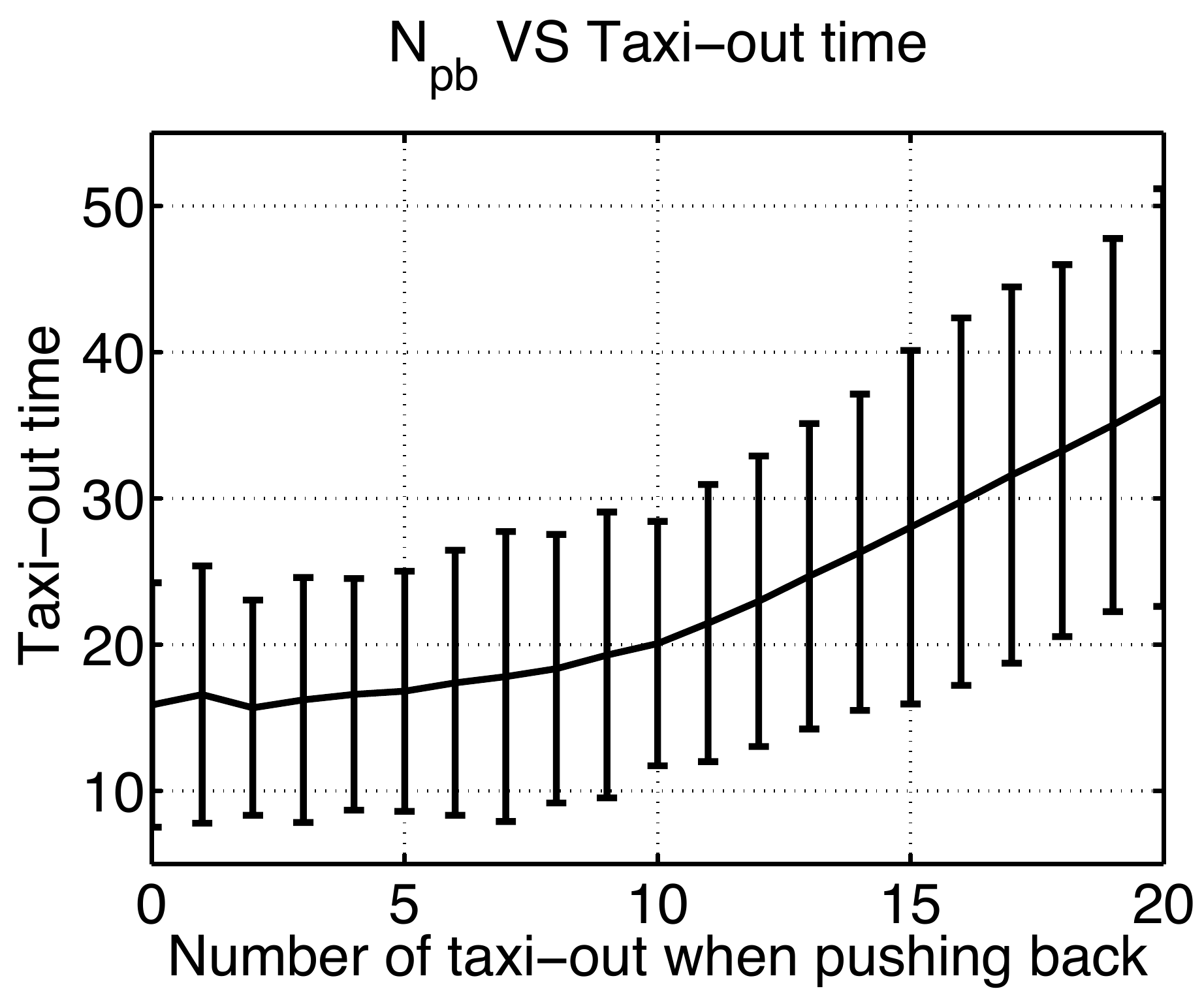}
	\caption{Taxi-out Time according to the Number of Taxi-out Aircraft When an Aircraft Pushes Back}
	\label{f:npbTO}
\end{figure}

\begin{figure}[!t]
	\centering
	\includegraphics[width=2.5in]{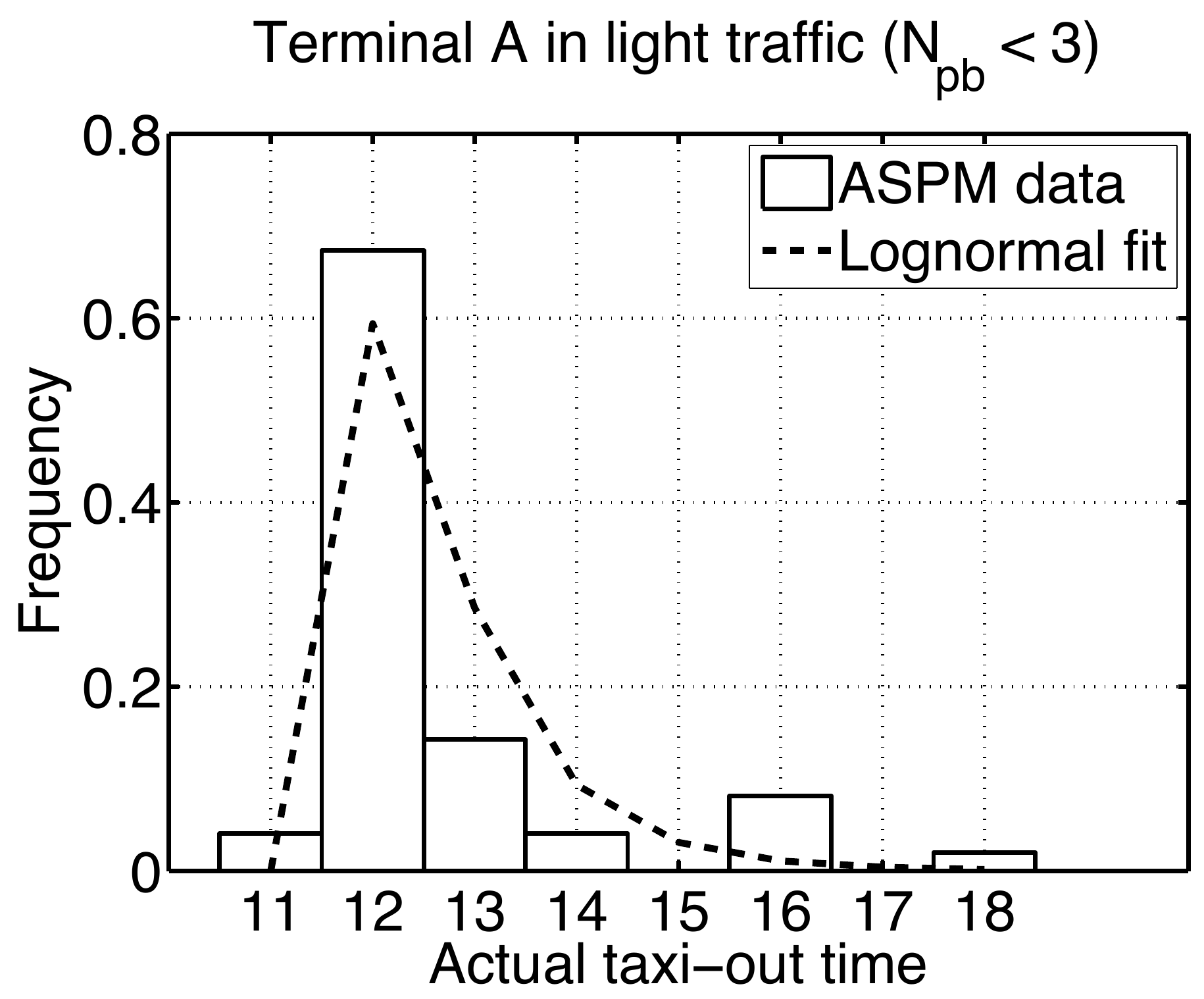}
	\caption{Taxi-out Time of Terminal A in Light Traffic}
	\label{f:termA}
\end{figure}

\begin{figure}[!t]
	\centering
	\includegraphics[width=2.5in]{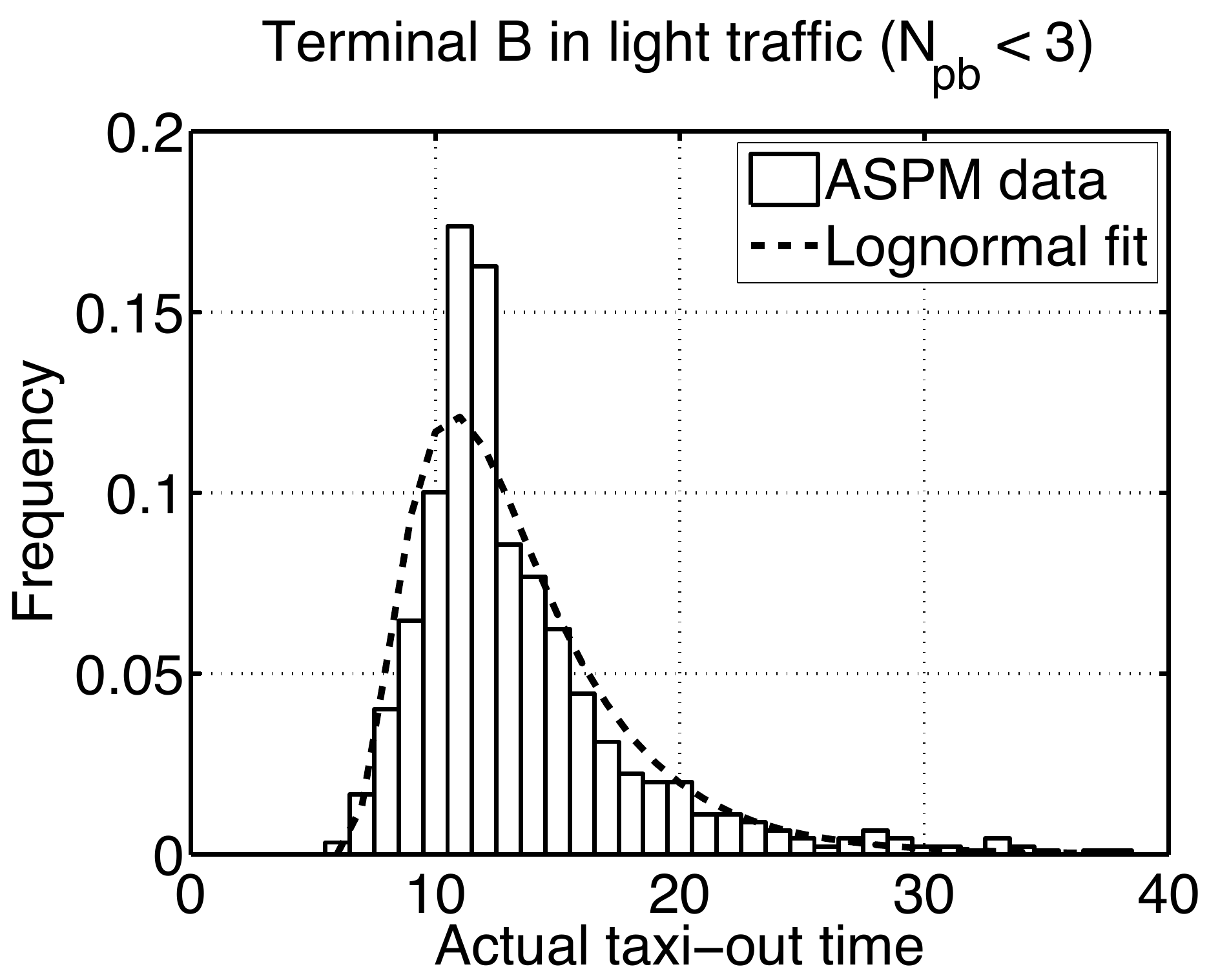}
	\caption{Taxi-out Time of Terminal B in Light Traffic}
	\label{f:termB}
\end{figure}

\begin{figure}[!t]
	\centering
	\includegraphics[width=2.5in]{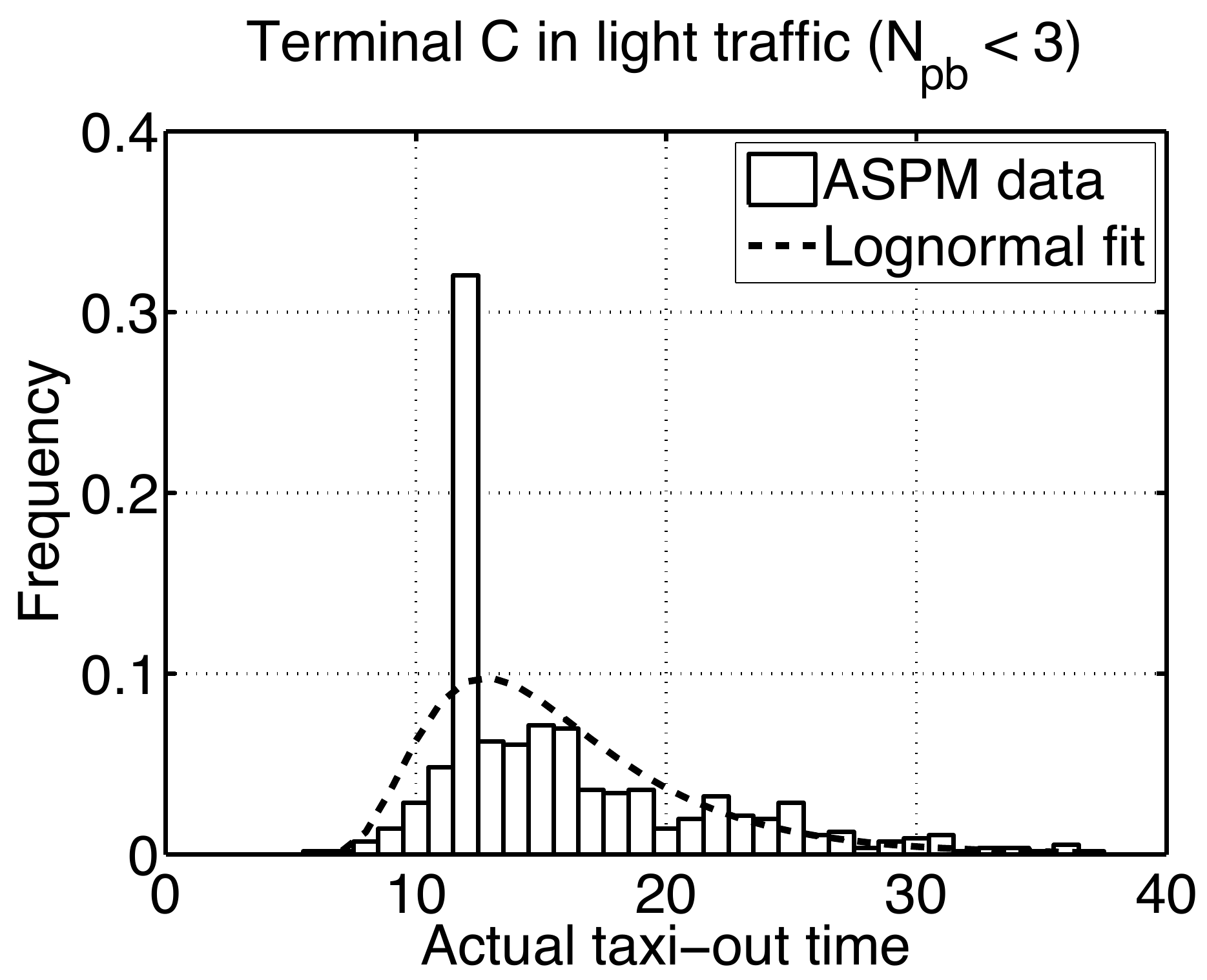}
	\caption{Taxi-out Time of Terminal C in Light Traffic}
	\label{f:termC}
\end{figure}

\begin{figure}[!t]
	\centering
	\includegraphics[width=2.5in]{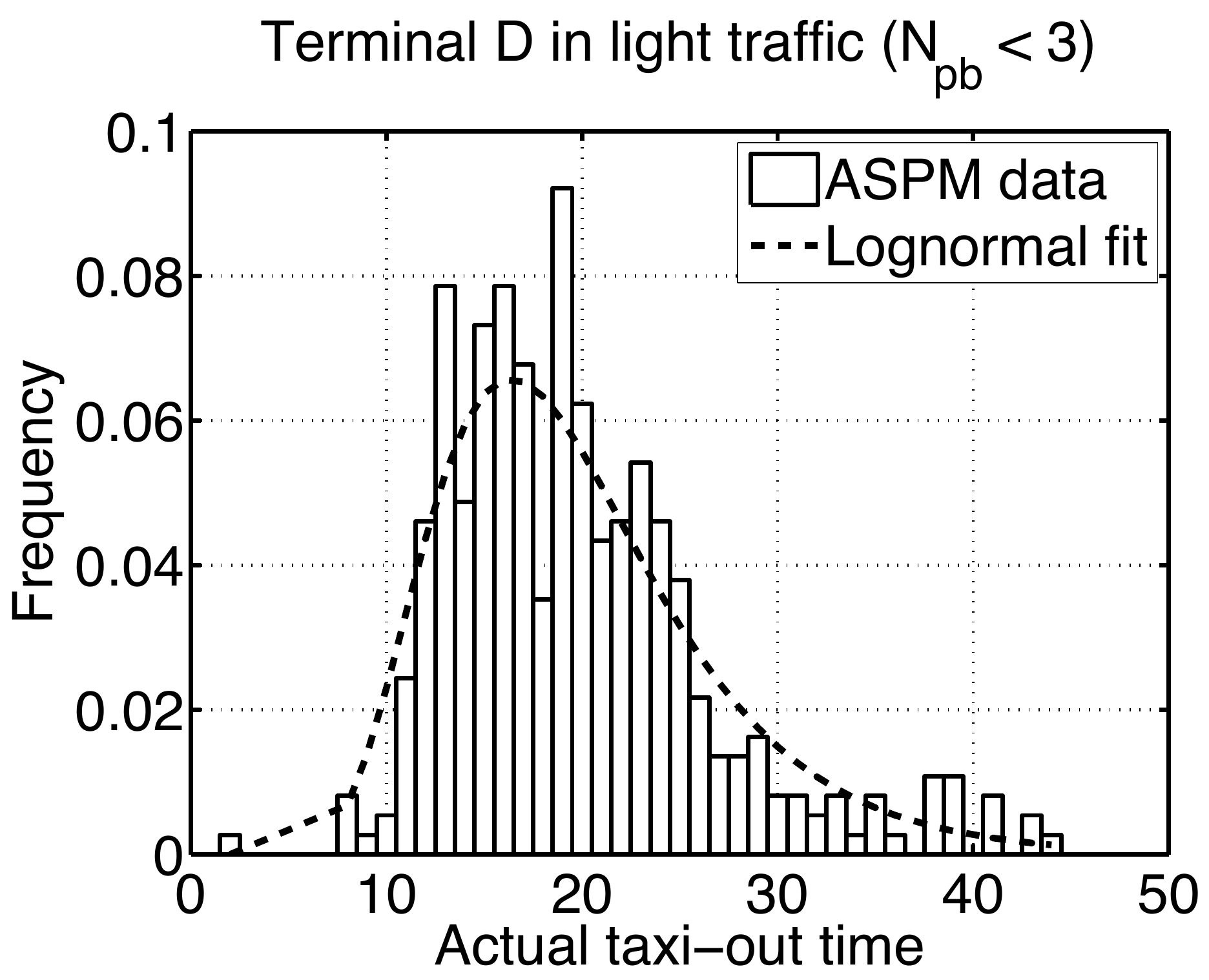}
	\caption{Taxi-out Time of Terminal D in Light Traffic}
	\label{f:termD}
\end{figure}

Taxi-out times in ASPM data are grouped by each terminal in LGA: Terminal A, B, C, and D. Airlines use mainly a single terminal in LGA; for instance, most flights of US Airways use terminal C. In order to get nominal taxi-out times, which are the taxiing times from a gate to a runway without a queuing delay on surface, taxi-out times are filtered by the number of taxi-out aircraft when an aircraft pushes back ($N_{pb}$). So, $N_{pb}$ indicates the number of departures that are on the way to the runway ahead of a pushing back aircraft. Fig.~\ref{f:npbTO} shows the means and the standard deviations of taxi-out time according to $N_{pb}$. The mean taxi-out time does not increase until $N_{pb}$ becomes 3. Hence, it is assumed that there is light traffic on the airport surface and taxi-out is unimpeded when $N_{pb} < 3$. A lognormal distribution is used to model the nominal taxi-out time, and Fig.~\ref{f:termA}-\ref{f:termD} show the taxi-out time of each terminal and their lognormal fits. Fig.~\ref{f:termA} and Fig.~\ref{f:termC} show an exceptionally high peak at 12 minutes and these terminals serve mostly small aircraft such as CRJ and ERJ. According to Simaiakis \cite{simaiakis2009}, this high peak is caused by a reporting issue of the ASPM data. Some airlines or aircraft do not participate to record their push-back, take-off, landing, and arrival times as known as OOOI times. For those flights, the taxi-out times are estimated using the median taxi-out time of the airport, which corresponds to the high peak. Due to the high peak, the lognormal model predicts the nominal taxi-out times of terminal A and C less accurately than terminal B and D.

\subsection{Model Validation}
\begin{figure}[!t]
	\centering
	\includegraphics[width=2.5in]{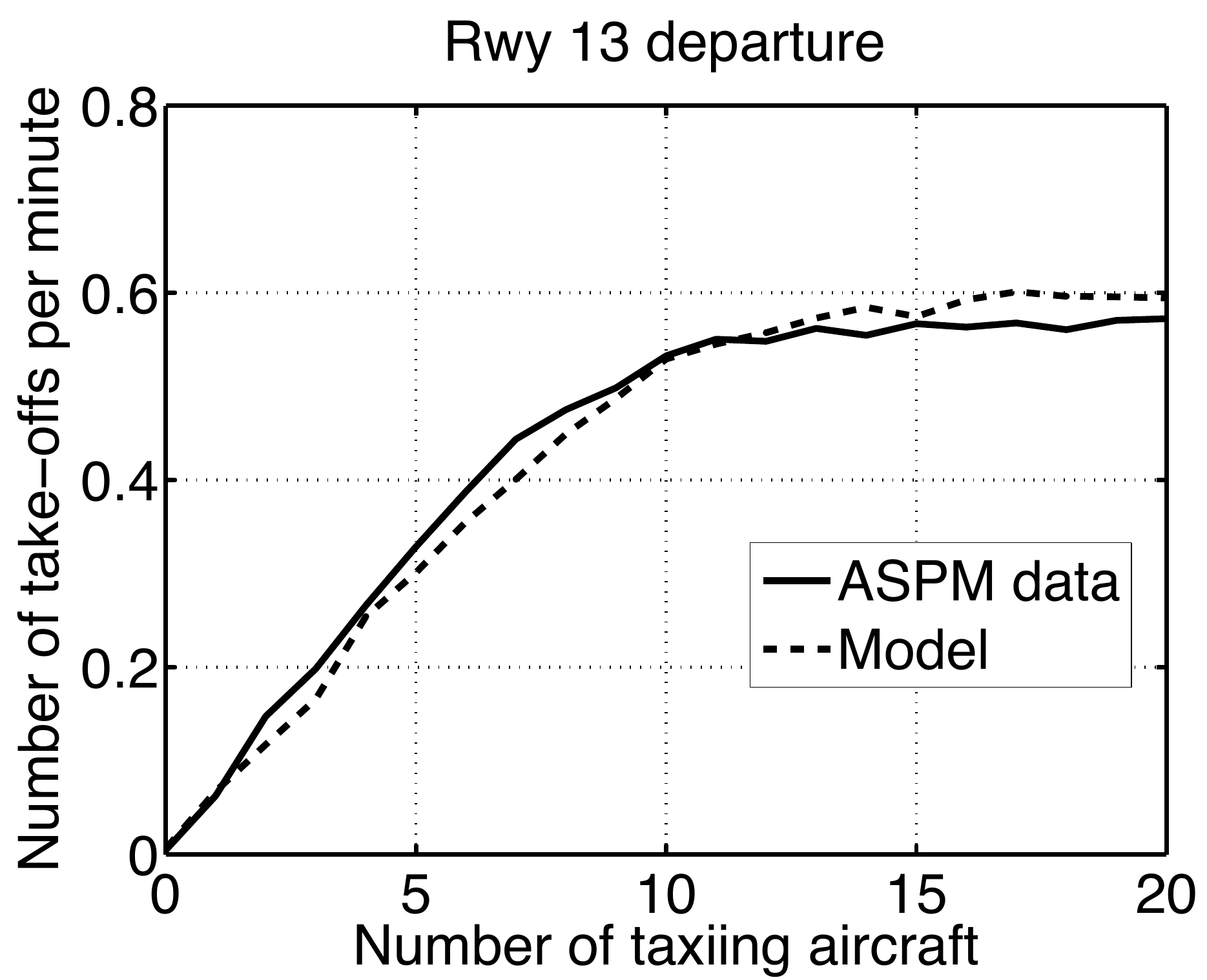}
	\caption{Model Validation: $T$ vs $N$}
	\label{f:valid1}
\end{figure}

\begin{figure}[!t]
	\centering
	\includegraphics[width=2.5in]{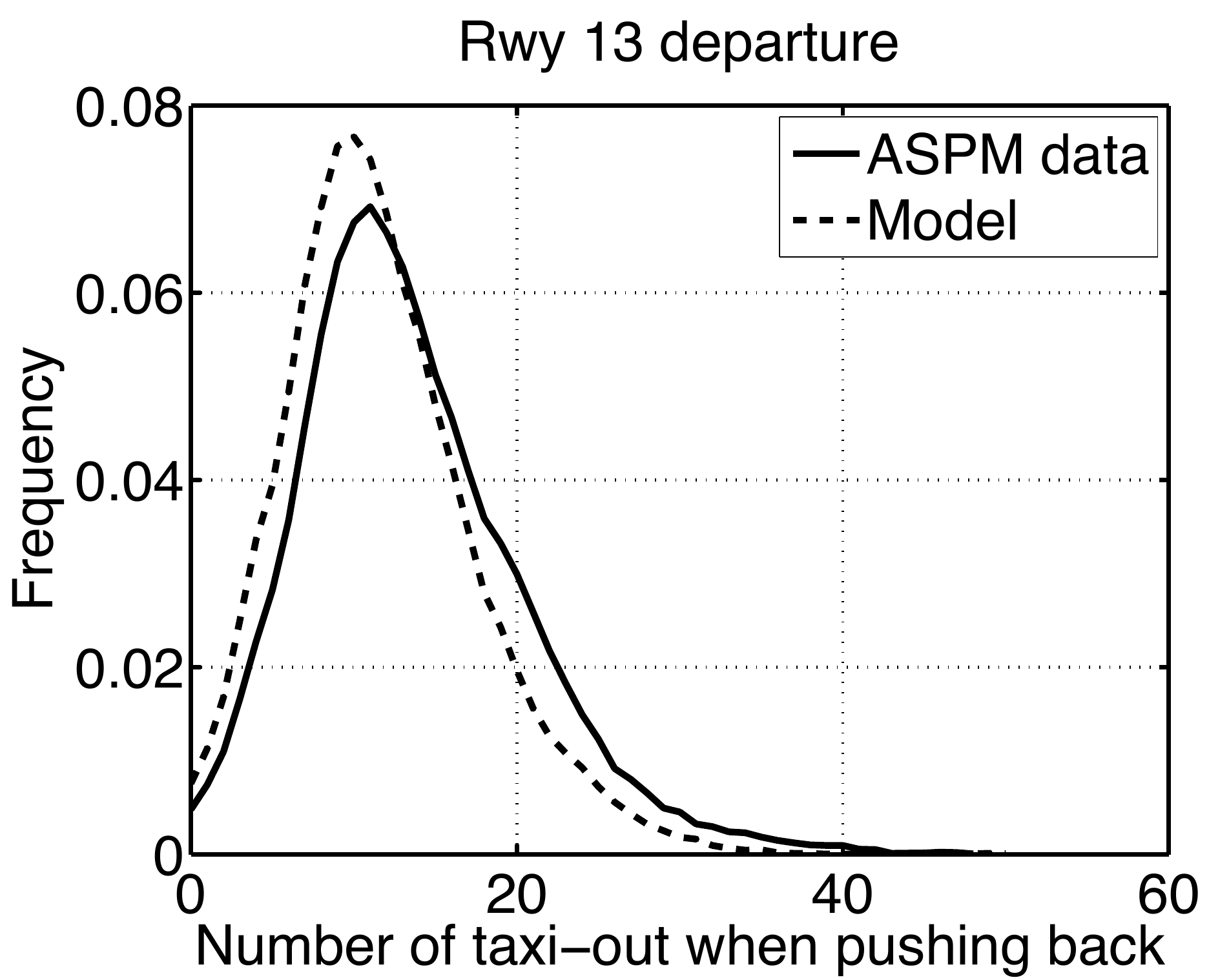}
	\caption{Model Validation: $N_{pb}$}
	\label{f:valid2}
\end{figure}

\begin{figure}[!t]
	\centering
	\includegraphics[width=2.5in]{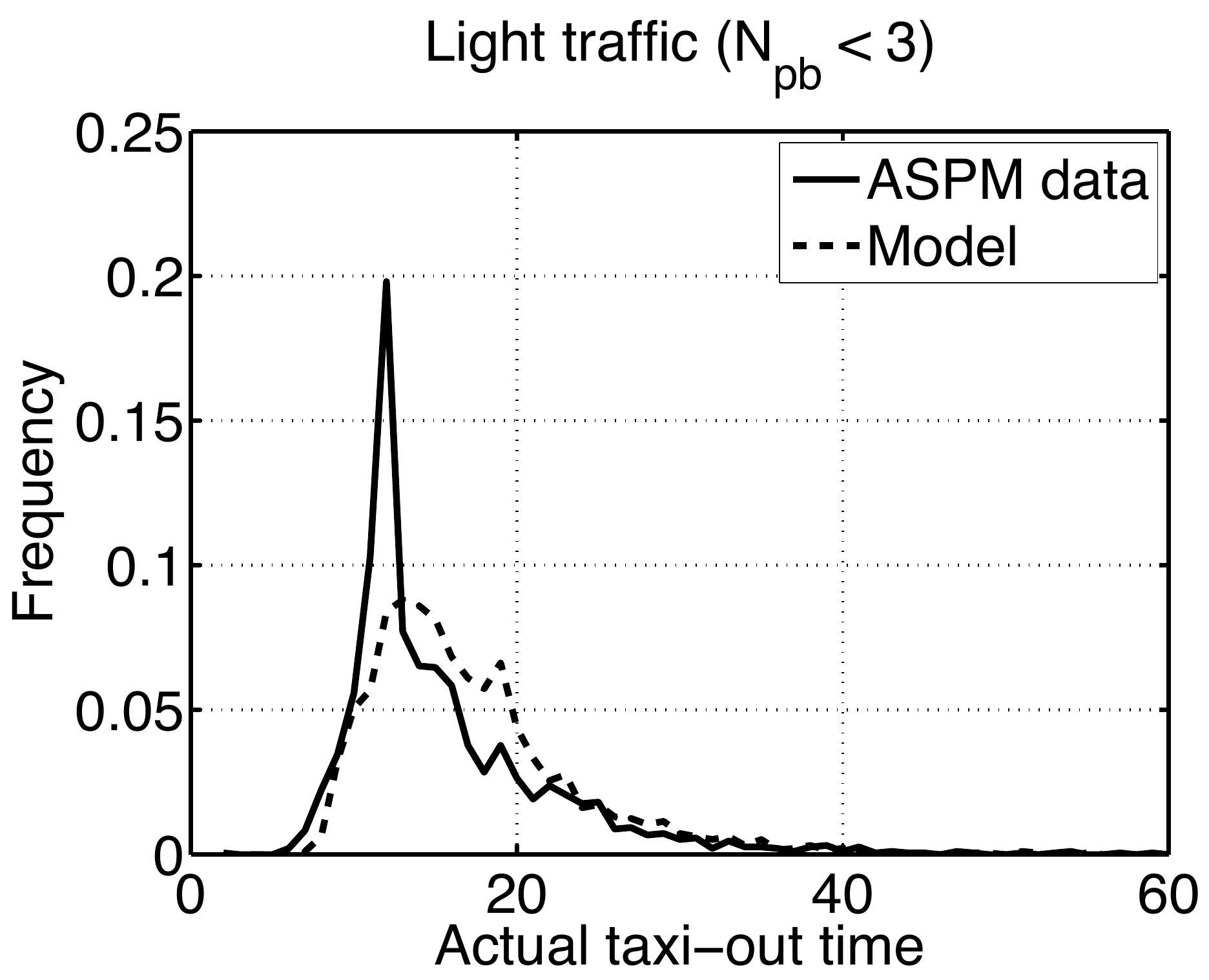}
	\caption{Model Validation: Taxi-out Time in Light Traffic}
	\label{f:light}
\end{figure}

\begin{figure}[!t]
	\centering
	\includegraphics[width=2.5in]{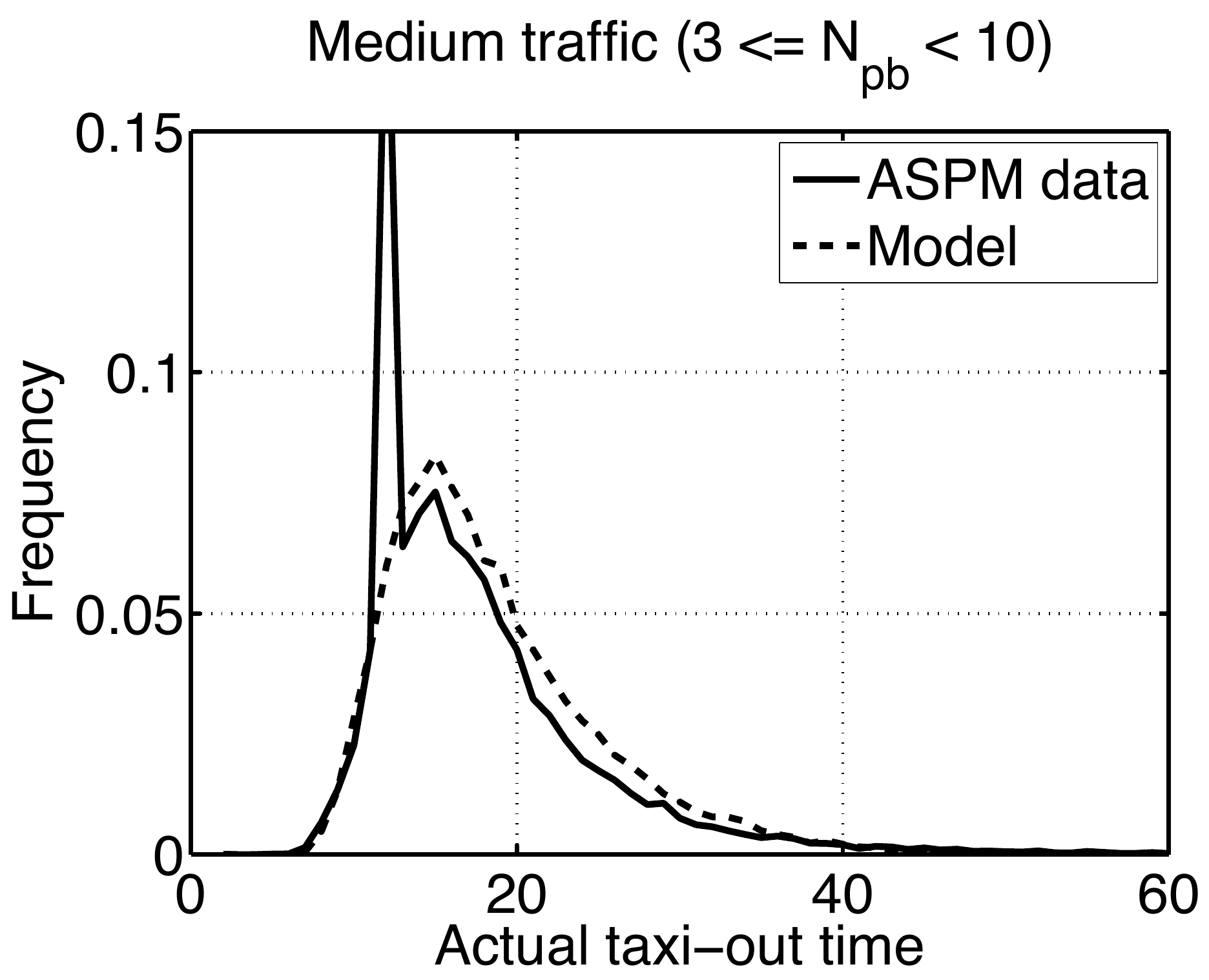}
	\caption{Model Validation: Taxi-out Time in Medium Traffic}
	\label{f:medium}
\end{figure}

\begin{figure}[!t]
	\centering
	\includegraphics[width=2.5in]{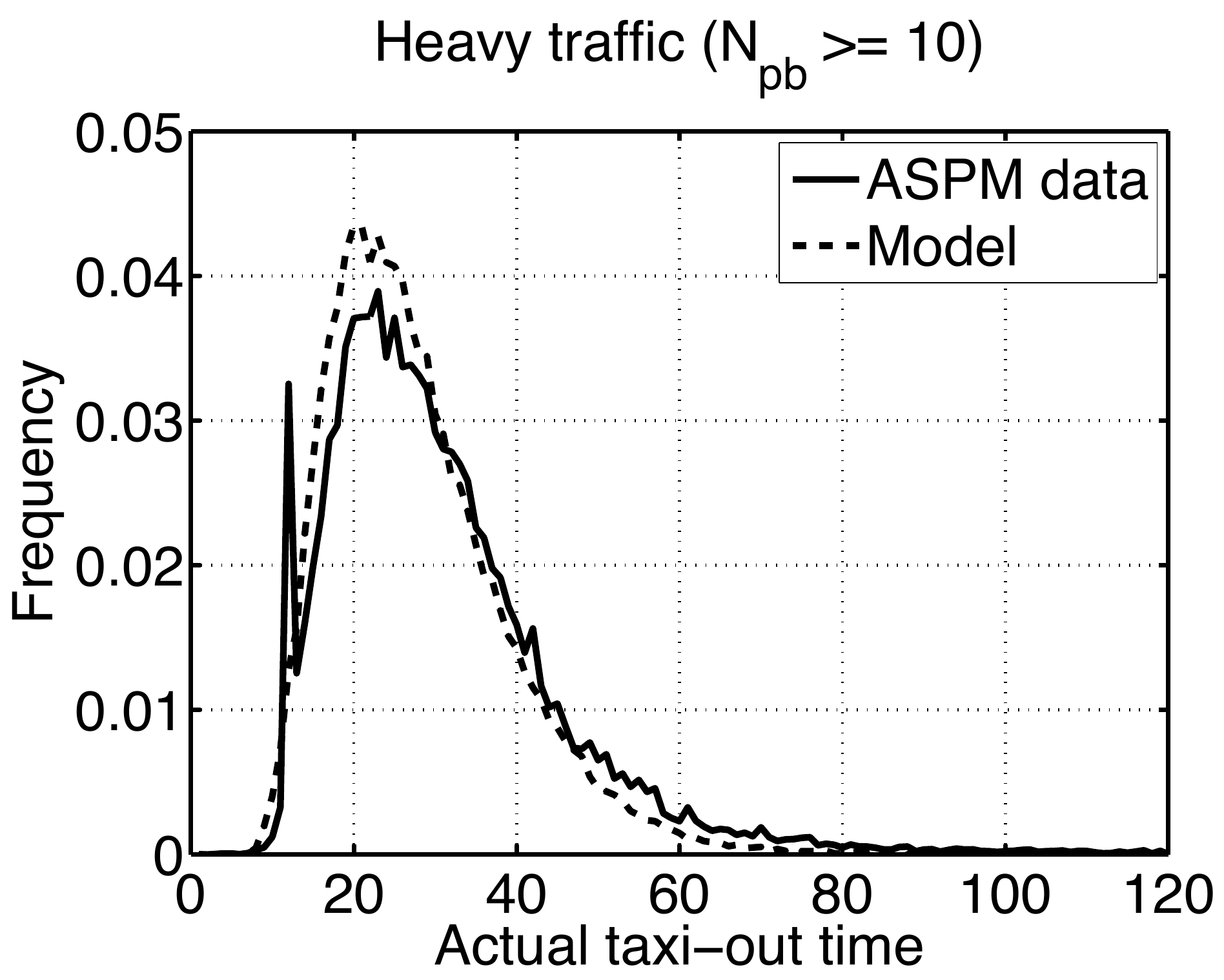}
	\caption{Model Validation: Taxi-out Time in Heavy Traffic}
	\label{f:heavy}
\end{figure}

The calibrated departure model is validated with departures on runway 13 in 2009. Fig.~\ref{f:valid1} shows the graph of $T$ vs $N$. The model reproduces the take-off rate well. Fig.~\ref{f:valid2} shows the distribution of $N_{pb}$. Fig.~\ref{f:light}-\ref{f:heavy} show the distribution of taxi-out times in light ($N_{pb}<3$), medium ($3<=N_{pb}<10$), and heavy ($N_{pb}>=10$) traffic. The model reproduces every traffic situation except for a high peak at 12 minutes, which is caused by the ASPM reporting issue along with the high peaks of Fig.~\ref{f:termA} and Fig.~\ref{f:termC}.
 
\section{Airport Gate Assignment}
\begin{figure}[!t]
	\centering
	\includegraphics[width=2.5in]{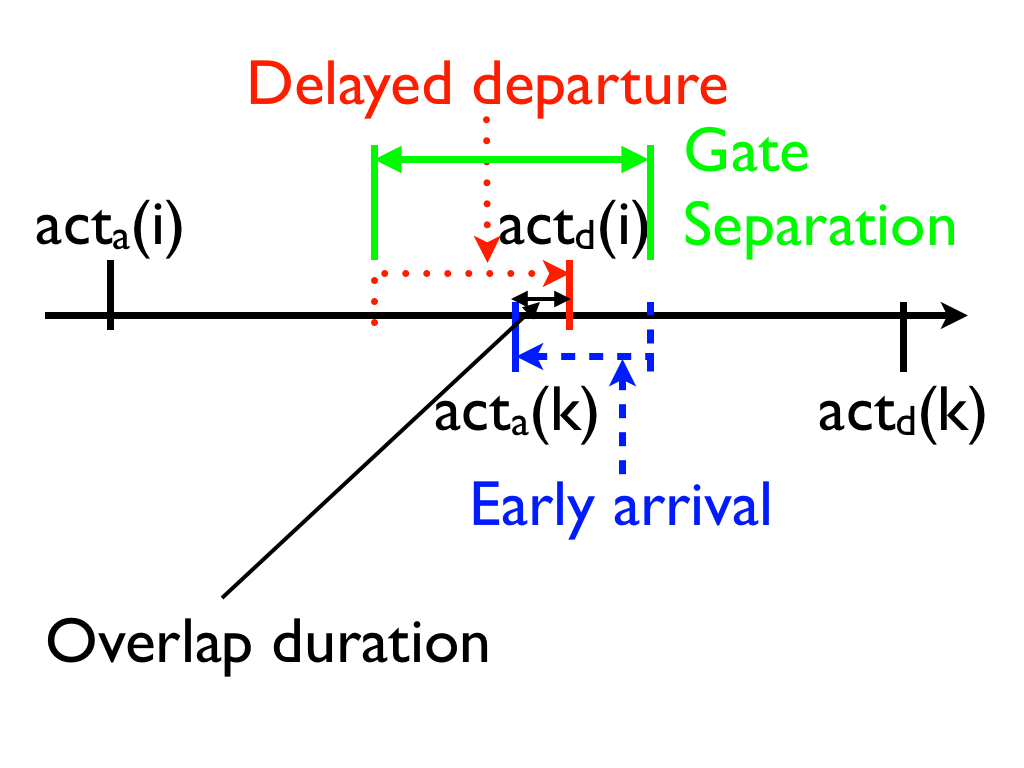}
	\caption{Gate Conflict and Overlap Duration}
	\label{f:schedule}
\end{figure}

If a departing aircraft is delayed at a gate and an arriving aircraft requests the gate, the arriving aircraft should wait until the departing aircraft pushes back and the gate is cleared, or be reassigned to another gate. This is known in the literature \cite{kim2011rga} as gate conflict and the duration of the overlap between the arrival time of the next aircraft and the departure time of the previous aircraft is a disturbance of gate assignment. Fig.~\ref{f:schedule} illustrates a gate conflict and the corresponding overlap duration, where $act_a(i)$ is the actual arrival time of flight $i$, $act_d(i)$ is the actual departure time of flight $i$, $act_a(k)$ is the actual arrival time of flight $k$, and $act_d(k)$ is the actual departure time of flight $k$.

\begin{figure}[!t]
	\centering
	\includegraphics[width=2.5in]{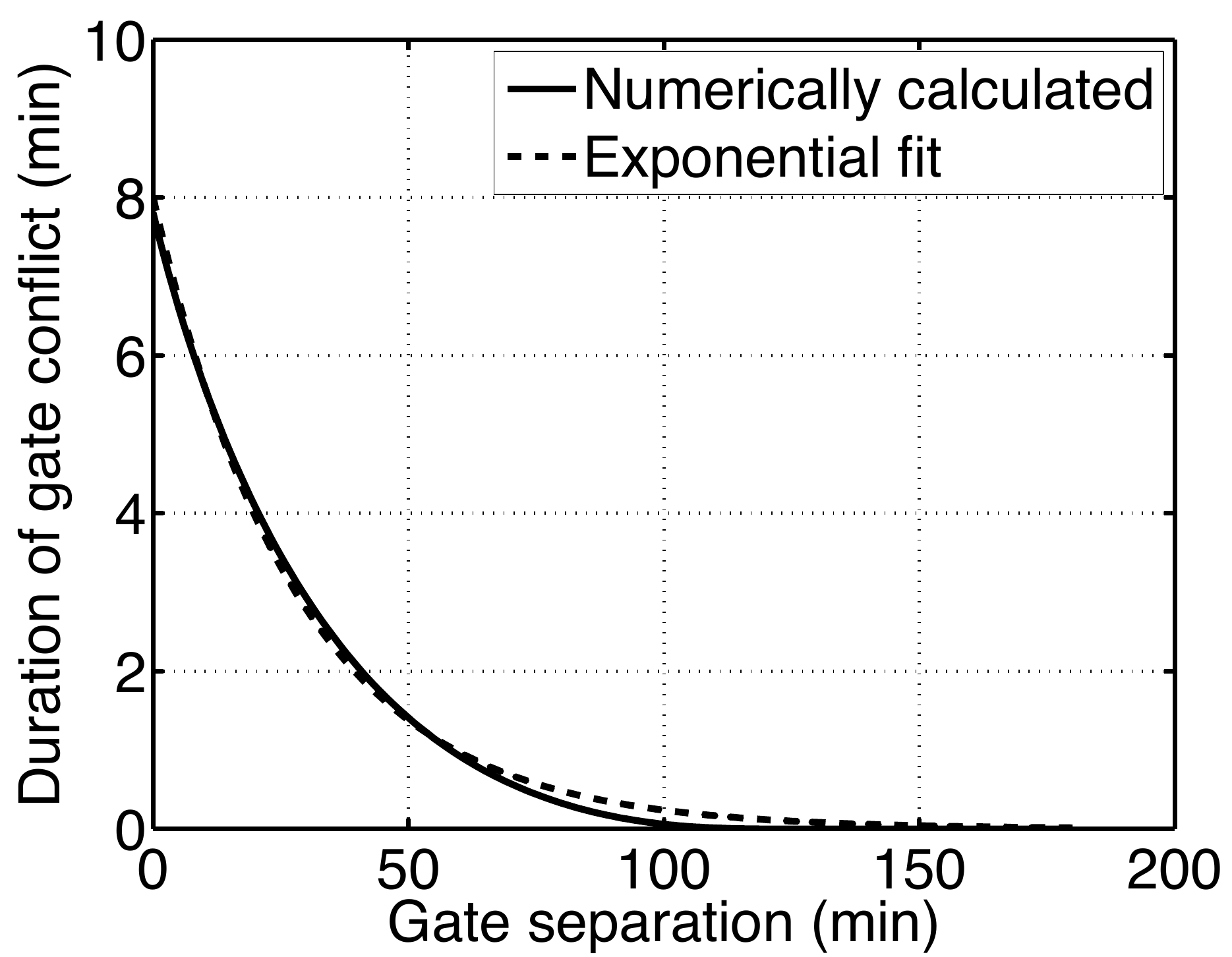}
	\caption{Expected Overlap Duration in a Function of Gate Separation}
	\label{f:overlap}
\end{figure}

Since arrival and departure delays are uncertain, the expected overlap duration at a given gate is calculated numerically with ASPM data. First, probability distributions of departure and arrival delays are generated for departure runway 13 of LGA in 2009. Then, the expected overlap duration is calculated $E[act_d(i)-act_a(k) | act_d(i)>act_a(k)]$ when the scheduled arrival time of flight $k$ is later than the scheduled departure time of flight $i$. Briefly speaking, the actual departure time ($act_d$) or the actual arrival time ($act_a$) is the sum of the scheduled departure or arrival time and departure/arrival delays. Because the scheduled times are fixed, $E[act_d(i)-act_a(k) | act_d(i)>act_a(k)]$ is a function of the departure delay of flight $i$ and the arrival delay of flight $k$. The probability distributions of the delays are derived from ASPM data. Details are given in \cite{kim2011rga}. The numerically calculated disturbance is fit to an exponential function $A \times B^x$, where $x$ is the gate separation. The variables $A$ and $B$ are 8 and 0.97 for LGA. Fig.~\ref{f:overlap} shows the numerically calculated value and the exponential fit. The exponential function is found to fit the numerically calculated value well. As seen in Fig.~\ref{f:overlap}, the expected overlap duration decreases as the gate separation increases. Note that the expected overlap duration is only about 8 minutes when gate separation is zero. The duration is surprisingly small because early departures and late arrivals occur frequently. More details on the delay distributions are available in \cite{kim2011rga}.

\begin{figure}[!t]
	\centering
	\includegraphics[width=2.5in]{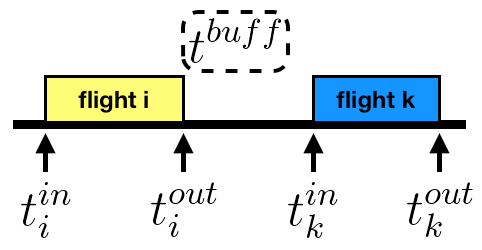}
	\caption{An Illustrative Example of Sufficient Gate Separation}
	\label{f:buffer}
\end{figure}

We write a formulation of the robust gate assignment problem as
\begin{align}
 \label{e:obj}
 &\text{Minimize} \sum_{i \in \mathcal{F}} \sum_{k \in \mathcal{F}, k > i} A\times B^{\text{sep}(i,k)} \sum_{j \in \mathcal{G}} \ x_{ij} \ x_{kj} \\
 &\nonumber \text{subject to} \\
 \label{e:gateconst}
 &\sum_{j \in \mathcal{G}} x_{ij} = 1, \ \forall i \in \mathcal{F} \\
 \label{e:feasconst}
 &(t^{\mbox{out}}_i - t^{\mbox{in}}_k + t^{\mbox{buff}}) (t^{\mbox{out}}_k - t^{\mbox{in}}_i + t^{\mbox{buff}}) \leq M (2 - x_{ij} - x_{kj}), \nonumber \\
 & i \neq k, \ \forall i, k \in \mathcal{F}, \ \forall j \in \mathcal{G} \\
 \label{e:decision}
 &x_{ij} \in \{0,1\}, \ \forall i \in \mathcal{F}, \ \forall j \in \mathcal{G} \\
 &\nonumber \text{where } x_{ij} = \left\{ 
  \begin{array}{rl}
   1 & \text{if } f_{i} \text{ is assigned to }g_{j} \\
   0 & \text{otherwise,}
   \end{array} \right.
\end{align}
where $\mathcal{F}$ is a set of flights, $\mathcal{G}$ is a set of gates, $\text{sep}(i,k)$ is the gate separation of aircraft $i$ and $k$, and $M$ is an arbitrarily large number. $x_{ij}$ is a decision variable of the optimization problem and has a value of 1 if aircraft $i$ is compatible with and is assigned to gate $j$, and 0 otherwise: some gates are not capable to serve certain types of aircraft. The first constraint (\ref{e:gateconst}) lets every aircraft be assigned to a single gate and the second constraint (\ref{e:feasconst}) enforces that gate separation between every pair of aircraft is larger than a certain minimum, which is $t^{\mbox{buff}}$. For instance, flight $i$ and $k$ are assigned to the same gate, and their gate occupancy times are shown in Fig.~\ref{f:buffer}. The gate separation of two flights, which is $t^{\mbox{in}}_k - t^{\mbox{out}}_i$, is greater than $t^{\mbox{buff}}$. Hence, the given assignment is feasible with respect to the second constraint (\ref{e:feasconst}). 

\begin{figure}[!t]
	\centering
	\includegraphics[width=2.5in]{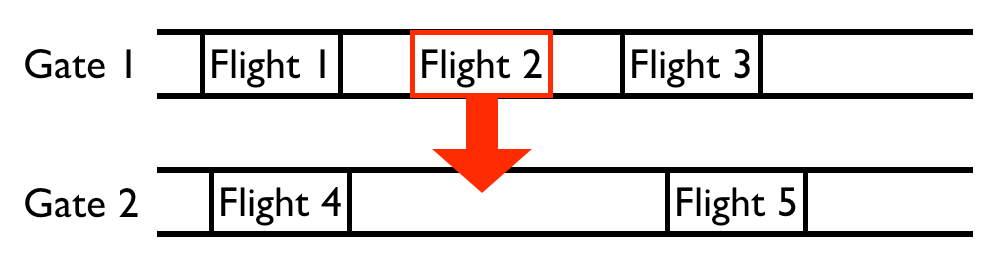}
	\caption{Insert Move}
	\label{f:insert}
\end{figure}

\begin{figure}[!t]
	\centering
	\includegraphics[width=2.5in]{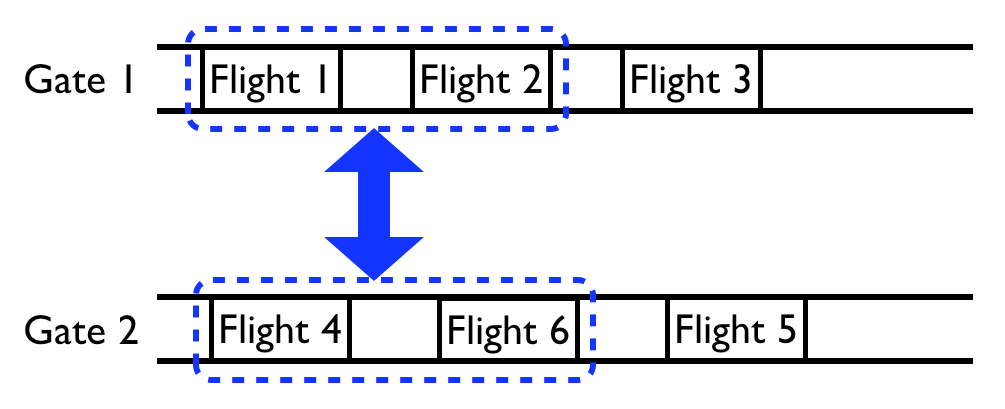}
	\caption{Interval Exchange Move}
	\label{f:exchange}
\end{figure}

The robust gate assignment problem is to minimize the total expected overlap duration. The problem is solved by Tabu Search (TS), which we have found to be an efficient method for solving gate assignment problems \cite{kim2012aga, kim2011rga}. TS utilizes the two types of neighborhood search moves shown in Fig.~\ref{f:insert}-\ref{f:exchange}. "Insert Move" switches the assigned gate of an aircraft to another gate and "Interval Exchange Move" swaps the assigned gate of a group of aircraft with that of another group of aircraft. Details are given in \cite{kim2012aga, kim2011rga}.

\section{Results}
The current gate assignment and a robust gate assignment are used to analyze the impact of gate assignment on gate holding. The current gate assignment for 5 days is obtained from a website (\url{www.flightstats.com}). The robust gate assignment is generated based on the schedule of the period because airport gates are assigned prior to the actual operation day.

The available flight schedules are separated into departures and arrivals, but an arrival and a departure sharing the same aircraft should be assigned to the same gate to eliminate any cost caused by towing the aircraft from one gate and to another. So, a departure is paired with an arrival by comparing the current gate assignment and the equipment type of each flight. It is frequently found in the current gate assignment that two arrivals use a gate consecutively and the gate is used for two consecutive departures. It means that the first arrival is towed to somewhere after it arrives, the second arrival arrives and departs, and then the first arrival is towed back to the gate for departure. In such a case, the corresponding arrival and departure are considered as a single arrival and a single departure that are not paired. Indeed, most flights are paired, and it removes any unnecessary towing. 

Each airline can use a subset of gates in LGA. For instance, US Airways uses most gates in terminal C. This airline-gate compatibility constrains the robust gate assignment problem. Most airlines use gates in a single terminal, but a few airlines have gates in multiple terminals. For instance, Delta Airlines also operates gates in terminal A and D. In 2013, Delta Airlines began to use gates in terminal C. The data used in this paper does not reflect this recent change in LGA. 

%

Table~\ref{t:separation} compares gate separations of the current gate assignment with those of the robust gate assignment. It is shown that the robust gate assignment makes the average gate separation longer than the current gate assignment does. Also, smaller standard deviation with the robust gate assignment indicates that the distribution of gate separation is less dispersed.

\begin{table}[!t]
	\caption{Comparison of Gate Separations}
	\label{t:separation}
	\centering
	\begin{tabular}{|c|c|c|}
		\hline
		& Current Gate Assignment & Robust Gate Assignment \\ \hline
		Mean Gate Separation & 94 min & 98.1 min \\
		Std Gate Separation & 155.7 min & 123.3 min \\ \hline
	\end{tabular}
\end{table}

\subsection{Simulation Model}
\begin{figure}[!t]
	\centering
	\includegraphics[width=2.5in]{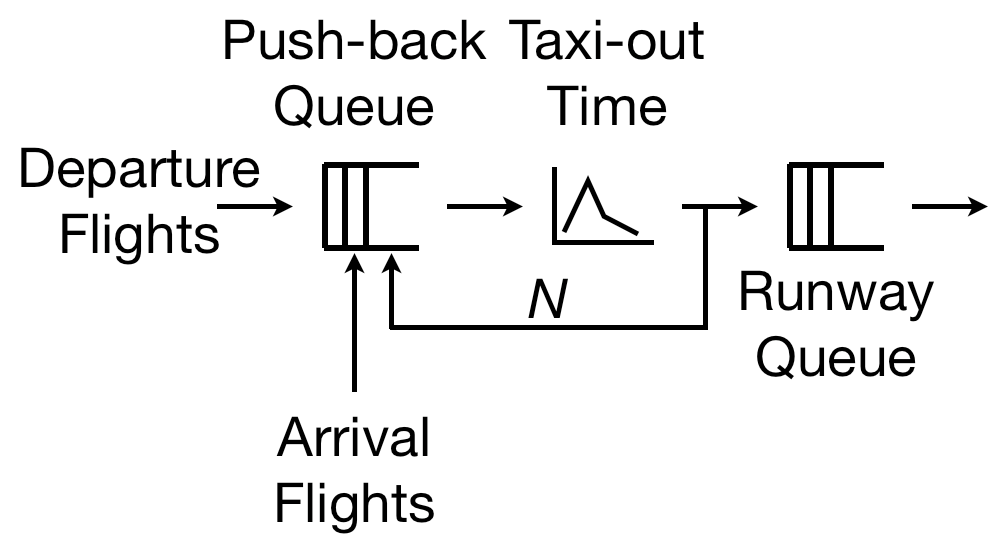}
	\caption{Simulation Structure}
	\label{f:model}
\end{figure}

The current gate assignment and the robust gate assignment are simulated using the airport departure model. The simulation structure is given in Fig.~\ref{f:model}. When a departure is ready to push back, it enters the push-back queue. A push-back is cleared FCFS, but if an arrival requests an occupied gate (gate conflict), the departure occupying the gate is cleared with the highest priority. When gate holding is active, push-back is not cleared until $N$ is below $N^*$. After the push-back, a taxi-out time is randomly generated according to the departure terminal and the aircraft enters the runway queue. From the runway queue, a take-off is cleared FCFS.

The simulation takes the actual departure and arrival times of the selected period as inputs. Note that gates are assigned based on the scheduled departure and arrival times. All the arrivals reach gates at actual arrival times and all the departures enter the push-back queue at actual departure times. The simulation runs 15 times and is averaged. 


\subsection{Relationship between Taxi-out Times, Gate-holding Times, and $N^*$}
\begin{figure}[!t]
	\centering
	\includegraphics[width=2.5in]{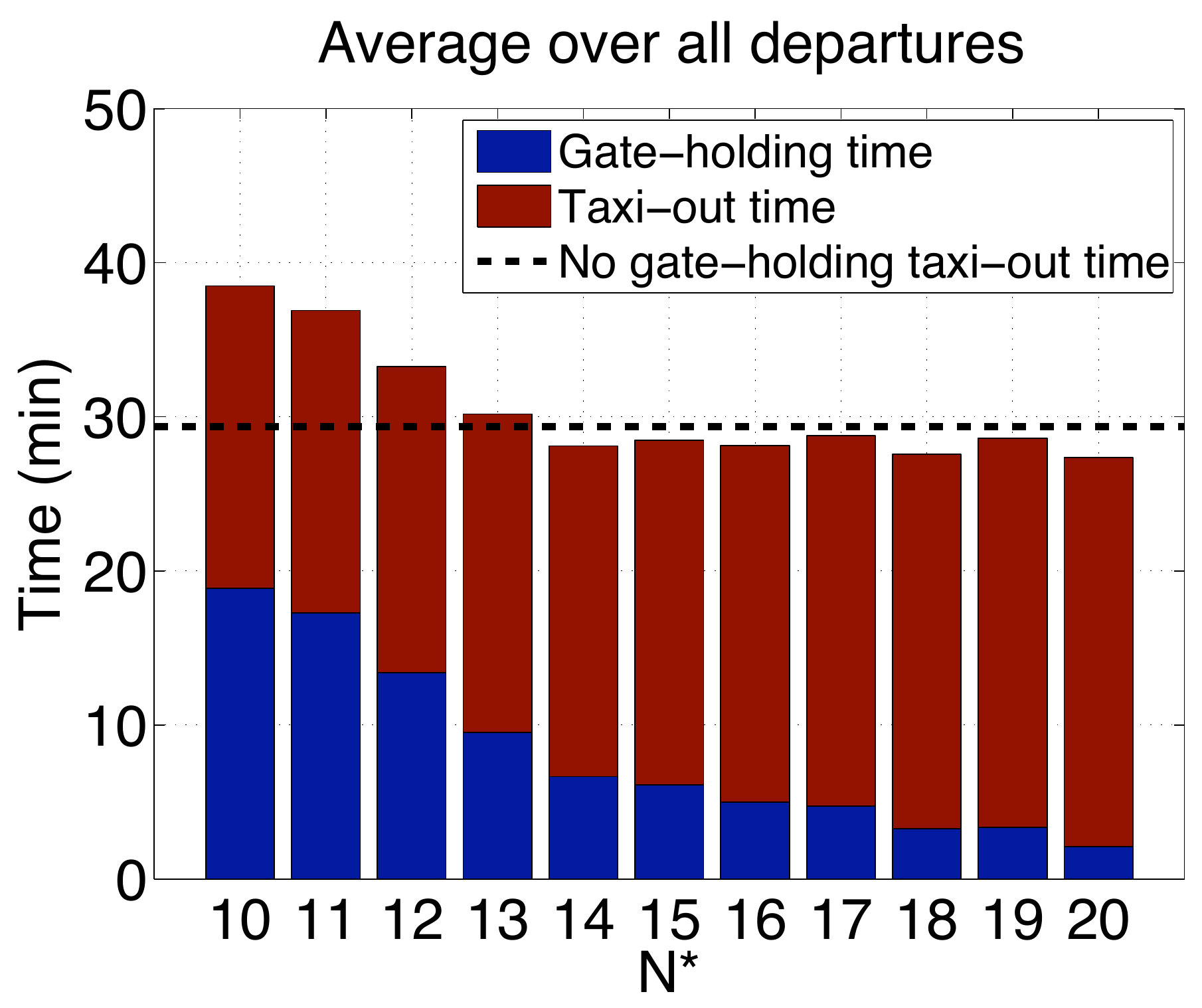}
	\caption{Gate-holding Times and Taxi-out Times for the Current Gate Assignment at LGA: The sums of gate-holding time and taxi-out time for $N^*$ equal to or greater than 14 are similar to the average taxi-out time without gate holding.}
	\label{f:simTaxiCurrent}
\end{figure}

\begin{figure}[!t]
	\centering
	\includegraphics[width=2.5in]{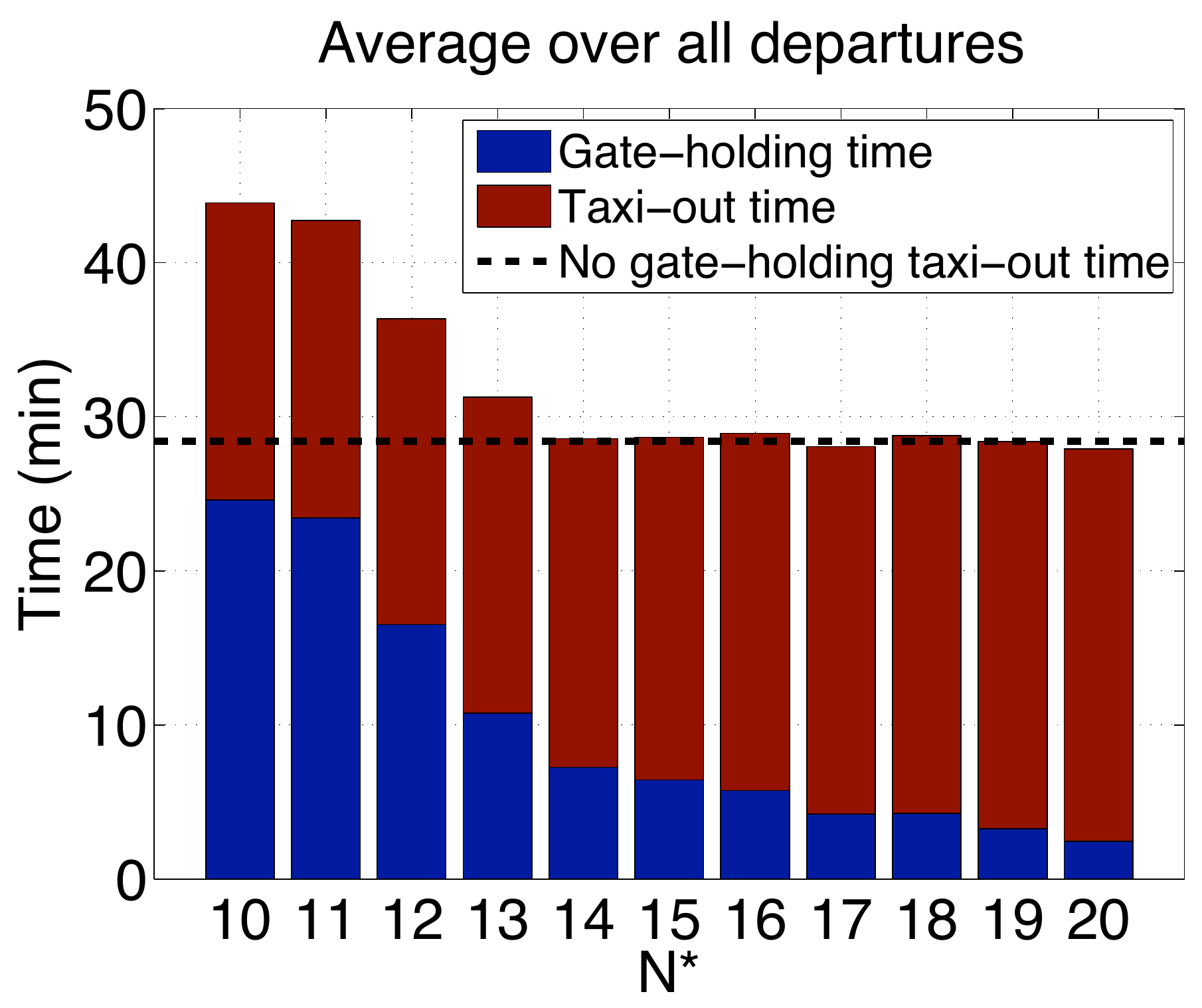}
	\caption{Gate-holding Times and Taxi-out Times for the Robust Gate Assignment at LGA: Like Fig.~\ref{f:simTaxiCurrent}, the sums of gate-holding time and taxi-out time for $N^*$ equal to or greater than 14 are similar to the average taxi-out time without gate holding.}
	\label{f:simTaxiRobust}
\end{figure}

Taxi-out times, gate-holding times, and $N^*$ are closely related. As $N^*$ increases, more departures are cleared to push back without metering. Hence, the airport surface becomes more congested and taxi-out times are likely to increase. On the other hand, when $N^*$ is low, many departures are held at the gates but taxiing aircraft can taxi out to the runway with less taxi delays. Fig.~\ref{f:simTaxiCurrent} and Fig.~\ref{f:simTaxiRobust} shows average gate-holding times and taxi-out times for the current gate assignment and the robust gate assignment with $N^*$ varying from 10 to 20. The gate-holding times are averaged over the whole set of departures, not only the gate-held departures. As predicted, gate-holding times decrease and taxi-out times increase as $N^*$ increases. Thus, the sum of gate-holding time and taxi-out time decreases as $N^*$ decreases, and remains constant for $N^*$ equal to or greater than 14. Note that the average taxi-out time without gate holding is similar to the sums of gate-holding times and taxi-out times for $N^*$ equal to or greater than 14. That means some taxi-out times are transferred to gate-holding times by gate holding. Then, gate-held departures can stay at gates without turning on their engines, and fuel consumption and emissions are reduced.

\subsection{Impact of Gate Assignment on Gate Holding}
\begin{table*}[!t]
	\caption{Impact of Gate Assignment on Gate Holding at LGA}
	\label{t:impact}
	\centering
	\begin{tabular}{|c|c|c|c|c|}
		\hline
		& \multicolumn{2}{|c|}{Current Gate Assignment} & \multicolumn{2}{|c|}{Robust Gate Assignment} \\ \hline
		& No Gate Holding & Gate Holding ($N^* = 14$) & No Gate Holding & Gate Holding ($N^* = 14$) \\ \hline
		Number of Gate Conflicts & 59 & 215.1 & 12.9 & 107.6 \\
		Number of Gate-held Departures & 0 & 1267.7 & 0 & 1315.1 \\
		Mean Gate-holding Times & 0 min & 12.6 min & 0 min & 13.3 min \\
		Mean Taxi-out Times & 29.4 min & 21.5 min & 28.4 min & 21.3 min \\ \hline
	\end{tabular}
\end{table*}

We analyze the impact of gate assignment on gate holding. Table~\ref{t:impact} compares the impacts of two gate assignments on gate holding. From Figs.~\ref{f:simTaxiCurrent}-\ref{f:simTaxiRobust}, $N^*$ is set to 14. With the current gate assignment, gate holding increases the occurrences of gate conflict over 3 times compared to no gate holding, and more than half the departures (1267.7 out of 2409) are held at gates for 12.6 minutes on average. As a consequence, some taxi-out times are transferred to gate delays by gate holding. Note that the mean gate-holding time is averaged over gate-held departures (1267.7 departures) as opposed to Figs.~\ref{f:simTaxiCurrent}-\ref{f:simTaxiRobust}, and the mean taxi-out time is averaged over the whole departures (2409 departures). So, the reduction of taxi-out time from gate holding (7.9 minutes) is smaller than the average gate-holding time (12.6 minutes). With the robust gate assignment, 1315.1 flights out of 2409 departures are held at gates for 13.3 minutes on average. The robust gate assignment induces fewer gate conflicts than the current gate assignment, whether or not gate holding is used as given in Table~\ref{t:impact}. Fewer gate conflicts of the robust gate assignment are due to the longer mean gate separation as shown in Table~\ref{t:separation}. Specifically, the robust gate assignment reduces the occurrence of gate conflicts by 78\% with gate holding and 50\% without gate holding compared to the current gate assignment. This demonstrates that the robust gate assignment helps gate holding get benefits with fewer disturbances to the gate assignment. When gate holding is active, departures are released from gates (cleared to push back) prior to an optimum time if the gates are requested by arrivals, and early release is expected to happen more with the current gate assignment as indicated by the number of gate conflicts. Early gate-release can induce an increase of taxi-out times. However, the average taxi-out time with the current gate assignment and active gate holding is just 0.2 min longer than that with the robust gate assignment and active gate holding. This number is somewhat smaller than expected. A possible explanation is that most gate-held departures are released at optimum times without being constrained by gate conflict. As given in Table~\ref{t:separation} and Table~\ref{t:impact}, mean gate separation (94 min for the current gate assignment and 98.1 min for the robust gate assignment) is much longer than the mean gate-holding times (12.6 min for the current gate assignment and 13.3 min for the robust gate assignment). Also, the ratios of the number of gate conflicts to the number of gate-held departures, which are 0.17 for the current gate assignment and 0.08 for the robust gate assignment, support this explanation. However, as indicated by high standard deviations of gate separations in Table~\ref{t:separation}, some aircraft have short gate separation and the robust gate assignment is beneficial for these aircraft.

\subsection{Expansion of the Model to Another Airport}
The queuing model is applied to another airport, which is one of the busiest hub airports in the U.S. Two airlines dominate the traffic of the airport.\footnote{Names of the airport and airlines are withheld to protect the airlines' data.} Carrier A operates 71.2 \% of operations, and carrier B operates 16.6 \% of operations. There are 5 runways in the airport, and 2 of them accommodate departures most of the time. The most frequently used runways for departure in 2009 are given in Table~\ref{t:atl_conf}. The meteorological condition of the airport is categorized by Visual Flight Rules (VFR) and Instrument Flight Rules (IFR). When the airport is under IFR, the runway throughput is reduced and the airport surface becomes more congested. Hence, the benefit of gate holding is likely to increase under IFR.

\begin{table}[!t]
	\caption{Frequently Used Runways for Departure at a U.S. Hub Airport}
	\label{t:atl_conf}
	\centering
	\begin{tabular}{|c|c|c|}
		\hline
		Runways & \% of Pushbacks in VFR & \% of Pushbacks in IFR \\ \hline \hline
		26L, 27R & 40.4 \% & 8.5 \% \\ \hline
		8R, 9L & 26.6 \% & 12.8 \% \\ \hline
	\end{tabular}
\end{table}

\begin{figure}[!t]
	\centering
	\includegraphics[width=2.5in]{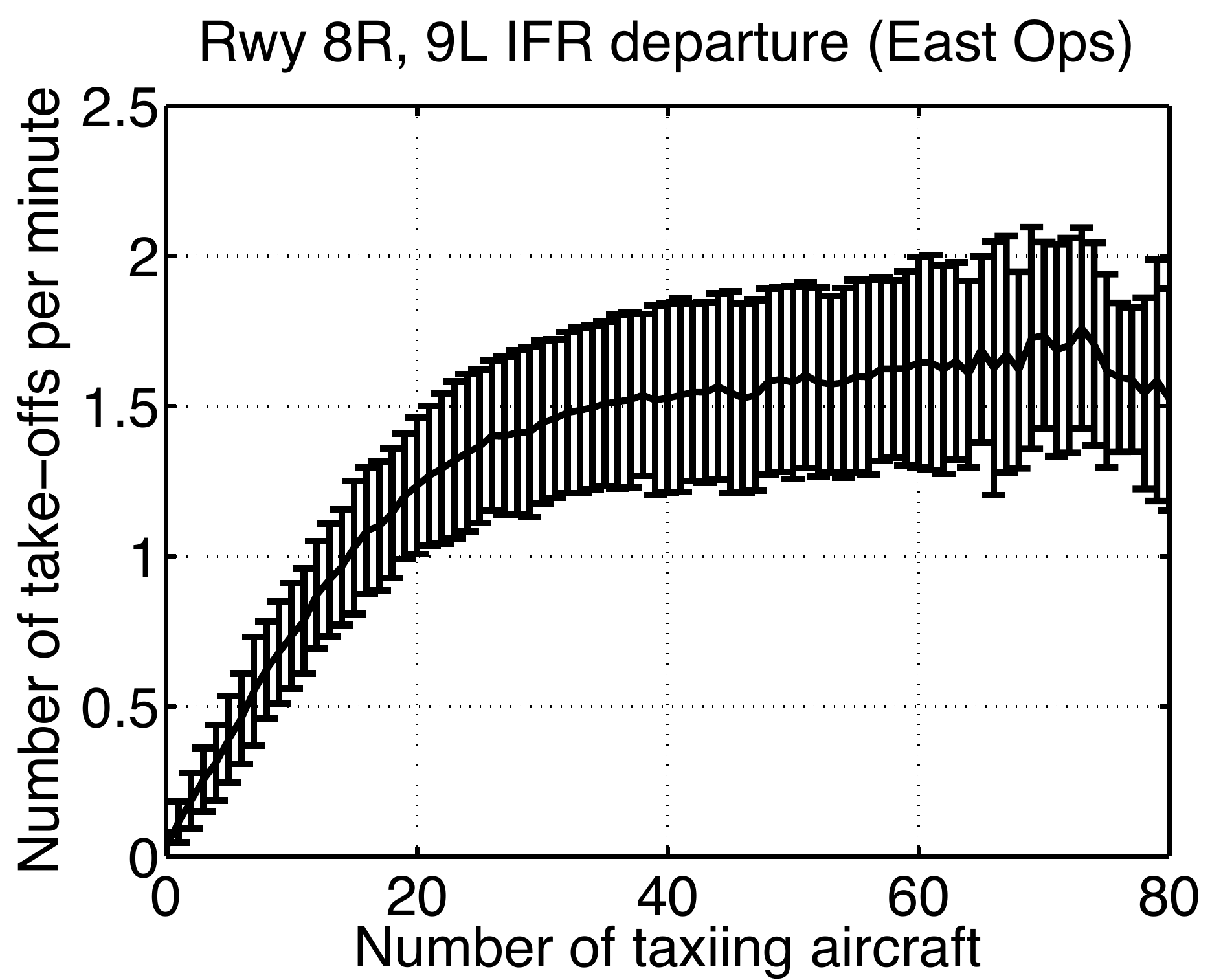}
	\caption{$T(t)$ as a Function of $N(t)$: The Vertical Bars Indicate the Standard Deviation of $T(t)$ for Each $N(t)$.}
	\label{f:atl_NT}
\end{figure}

\begin{figure}[!t]
	\centering
	\includegraphics[width=2.5in]{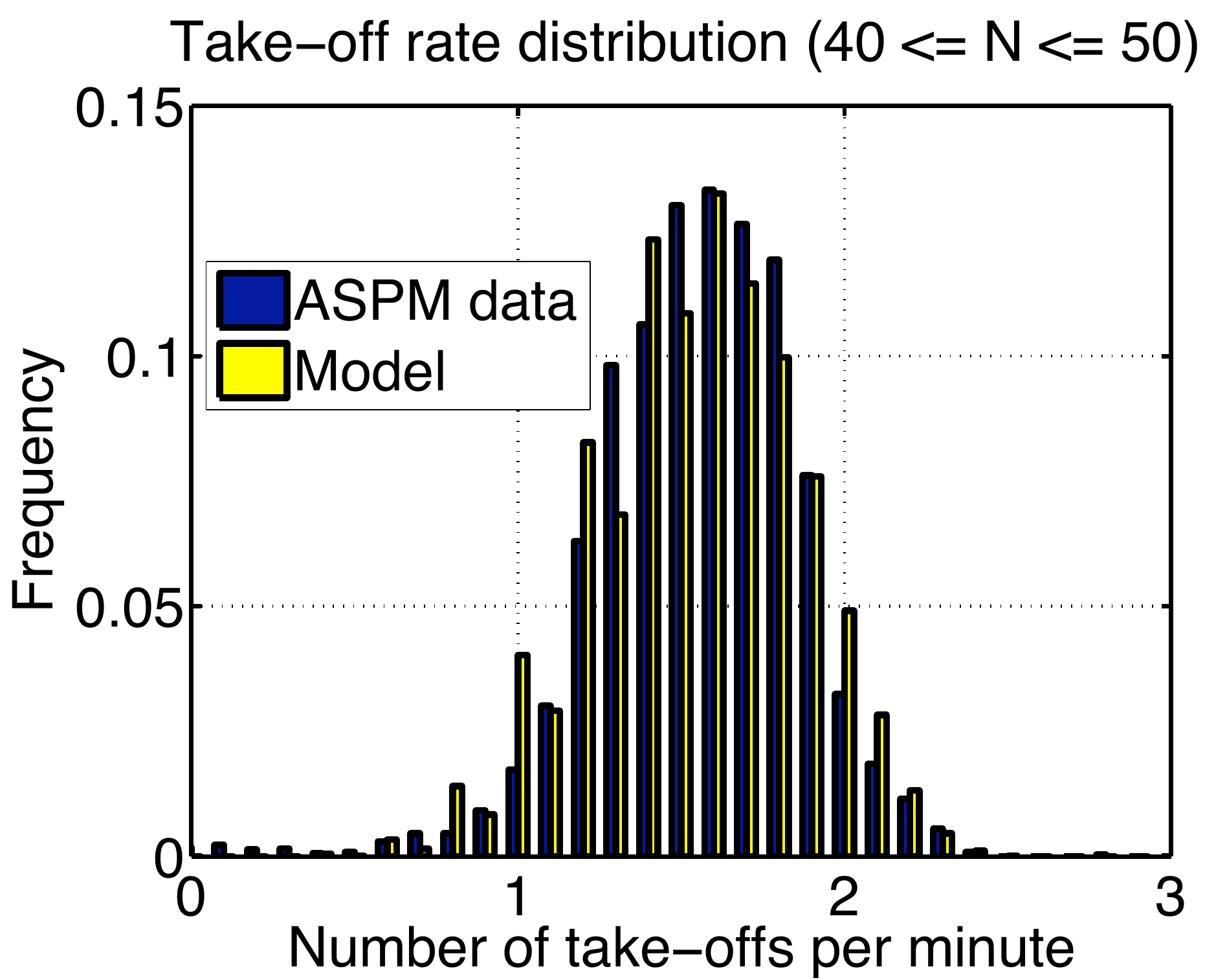}
	\caption{Take-off Rate Distribution}
	\label{f:atl_takeoff}
\end{figure}

The queuing model is calibrated with departures from 8R and 9L runways under IFR. The corresponding departures account for 12.8 \% of departures in 2009. The mean and standard deviation of the airport throughput $T(t)$ vs $N(t)$ is shown in Fig.~\ref{f:atl_NT}. Similar to Fig.~\ref{f:NT}, $T(t)$ increases with $N(t)$ and is saturated when $N(t)$ is larger than a certain number. It is shown that $T(t)$ drops when $N(t)$ is higher than 70. This drop might indicate gridlock and match that to ground transportation literatures \cite{daganzo1994cell, wong2002multi}. The take-off model is calibrated from $T(t)$ when $N(t)$ is in the range of [40, 50], and it is shown in Fig.~\ref{f:atl_takeoff}. 

\begin{figure}[!t]
	\centering
	\includegraphics[width=2.5in]{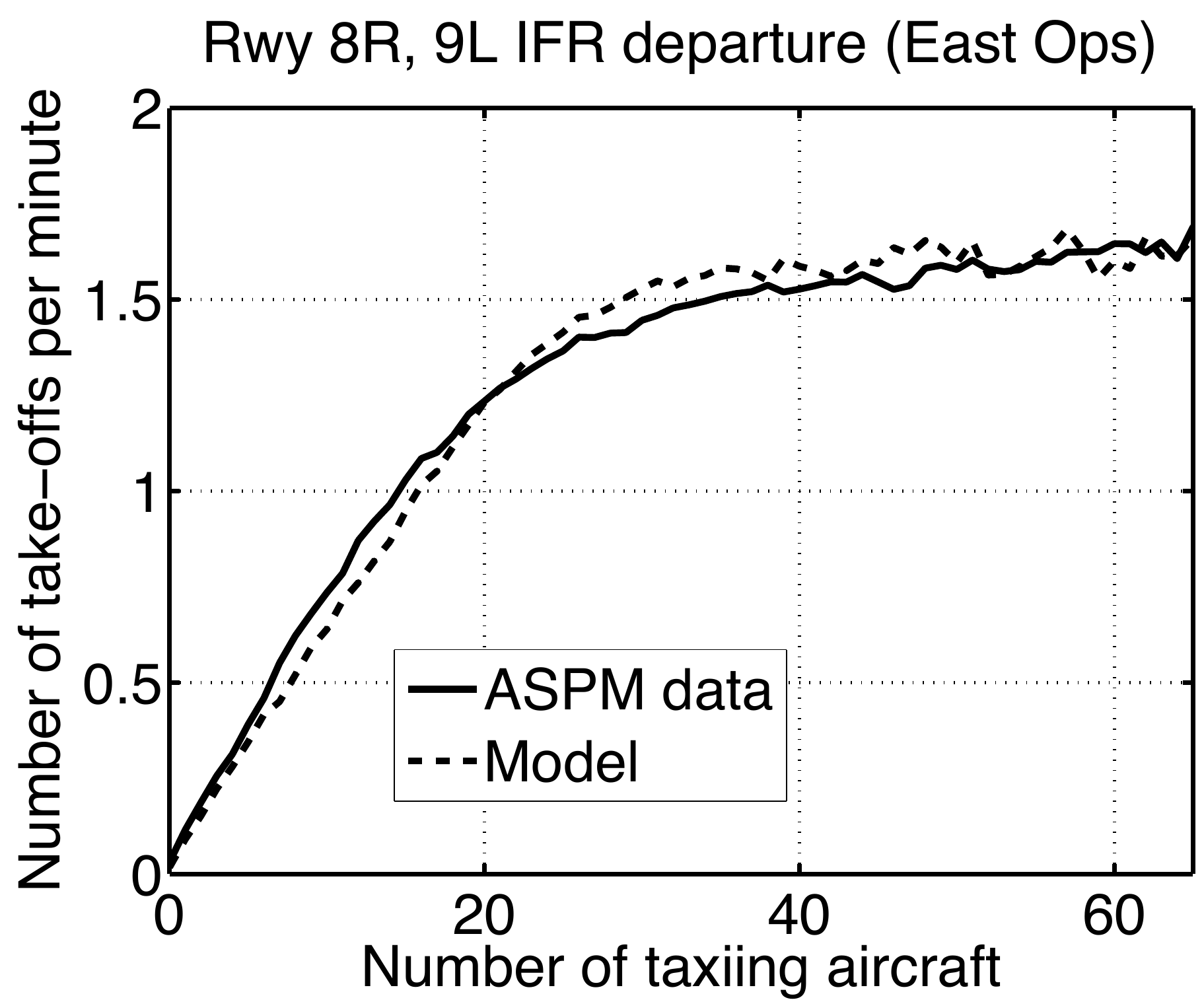}
	\caption{Model Validation: $T$ vs $N$}
	\label{f:atl_NTmodel}
\end{figure}

\begin{figure}[!t]
	\centering
	\includegraphics[width=2.5in]{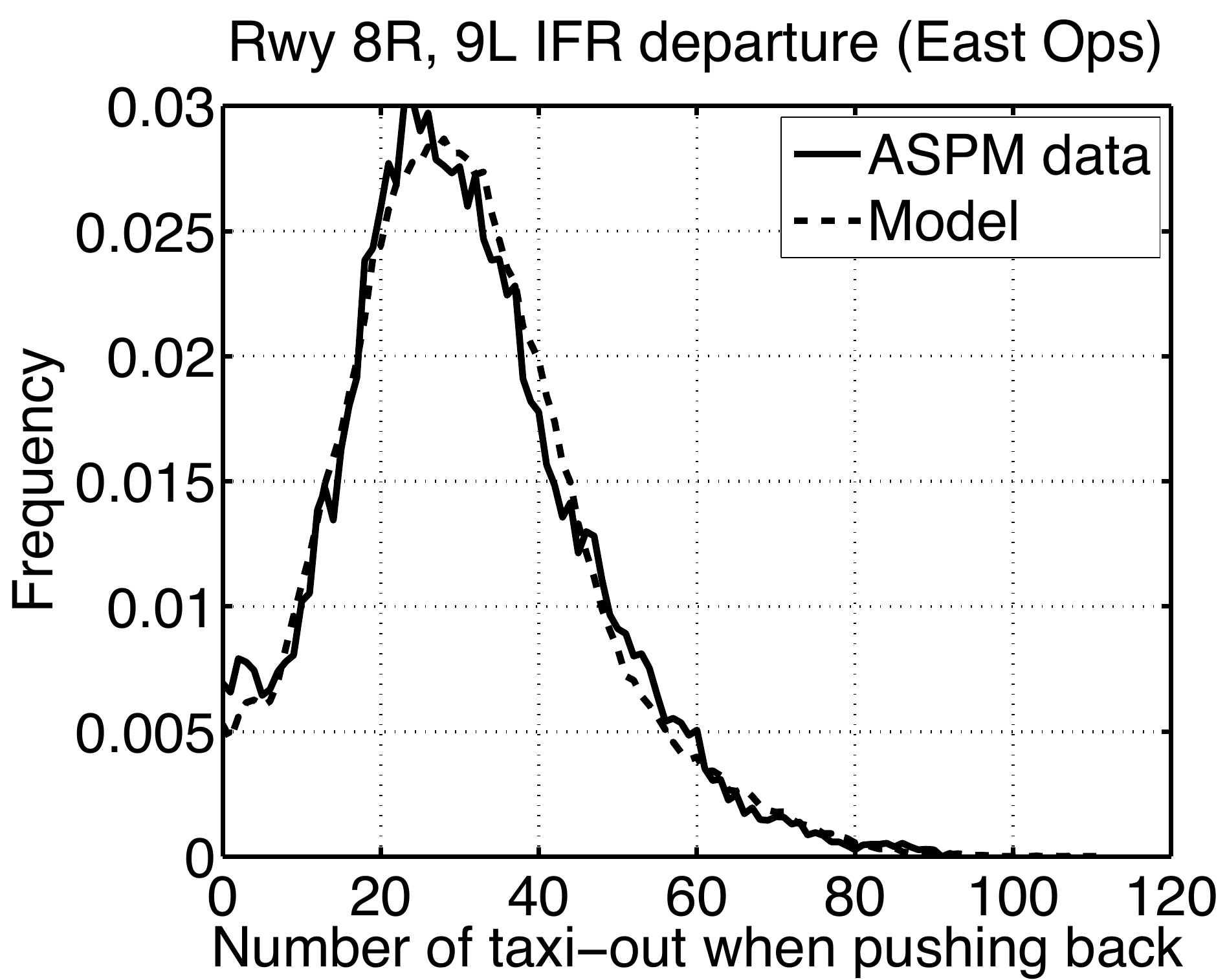}
	\caption{Model Validation: $N_{pb}$}
	\label{f:atl_Npb}
\end{figure}

The $T$ vs $N$ curve is given in Fig.~\ref{f:atl_NTmodel}. It is shown that the model reproduces the saturation of the departure throughput well. Also, the model simulates surface congestion as shown in Fig.~\ref{f:atl_Npb}. $N_{pb}$, which is the number of taxi-out aircraft when an aircraft pushes back, indicates how airport surface is congested when each aircraft leaves the gate. From Figs.~\ref{f:atl_NTmodel}-\ref{f:atl_Npb}, the model is successful to simulate departure operations in every traffic situation.

\begin{figure}[!t]
	\centering
	\includegraphics[width=2.5in]{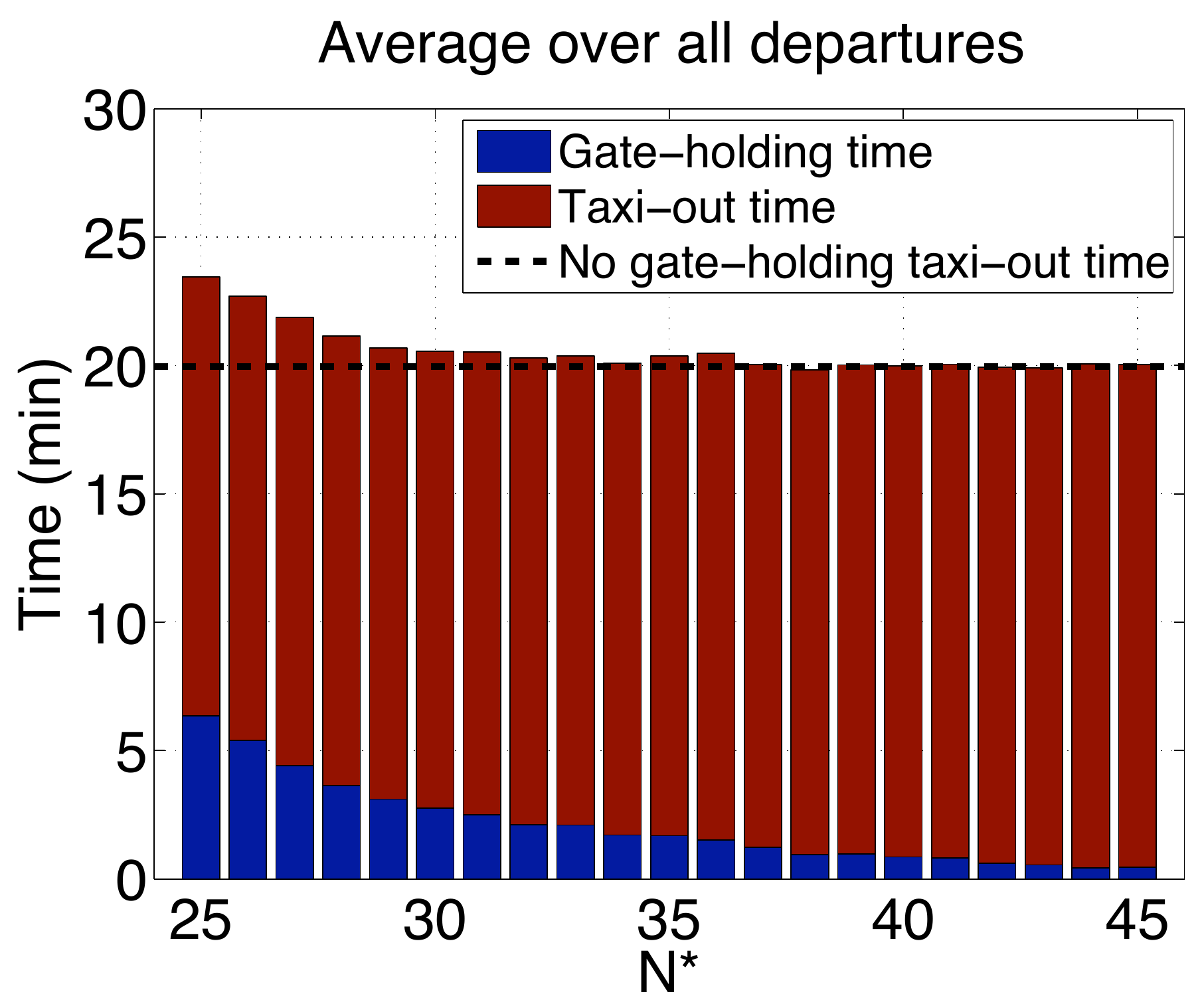}
	\caption{Gate-holding Times and Taxi-out Times for the Current Gate Assignment at a U.S. Hub Airport: The sums of gate-holding time and taxi-out time for $N^*$ equal to or greater than 14 are similar to the average taxi-out time without gate holding.}
	\label{f:atl_simTaxiCurrent}
\end{figure}

\begin{figure}[!t]
	\centering
	\includegraphics[width=2.5in]{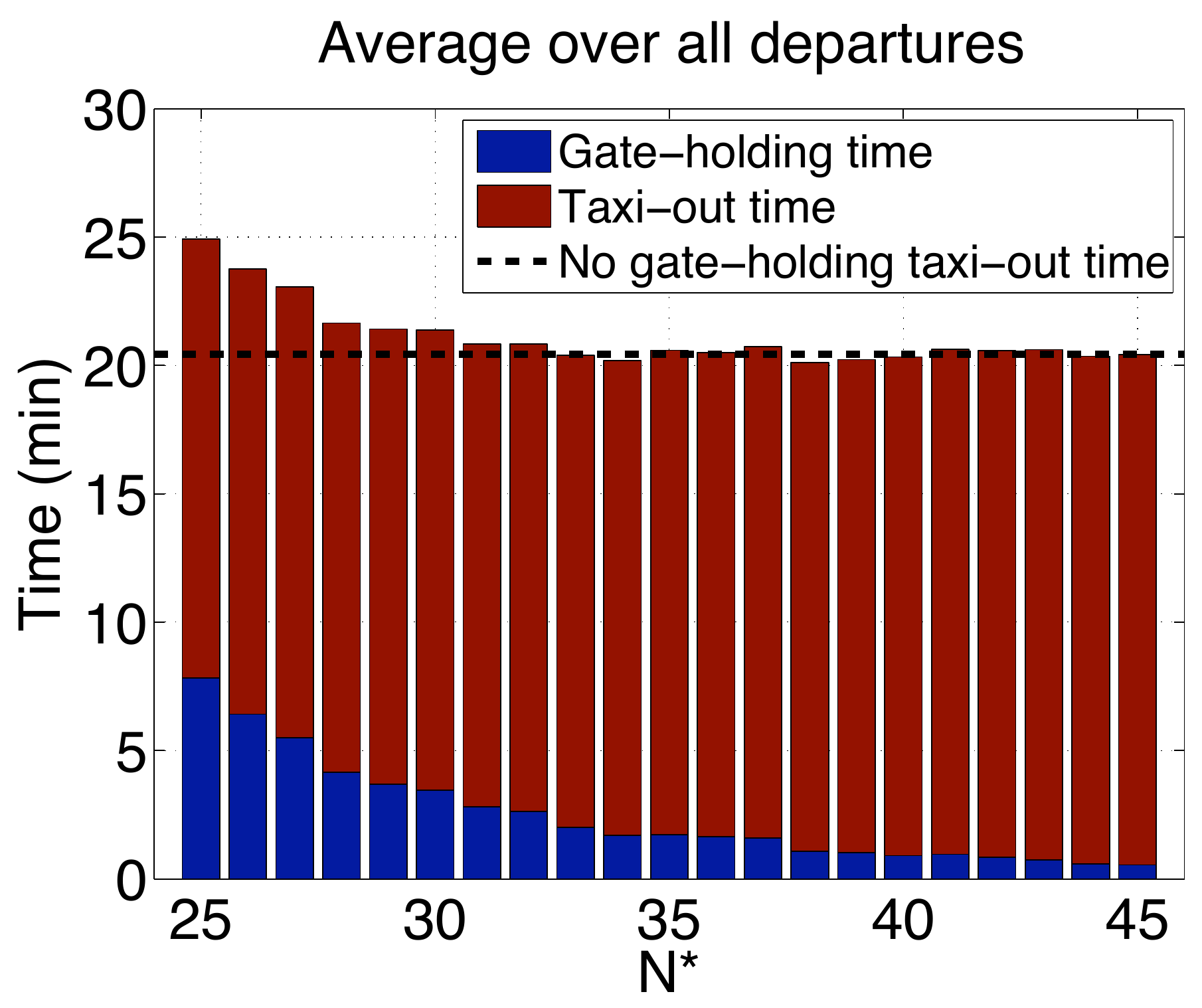}
	\caption{Gate-holding Times and Taxi-out Times for the Robust Gate Assignment at a U.S. Hub Airport: The sums of gate-holding time and taxi-out time for $N^*$ equal to or greater than 14 are similar to the average taxi-out time without gate holding.}
	\label{f:atl_simTaxiRobust}
\end{figure}

\begin{figure}[!t]
	\centering
	\includegraphics[width=2.5in]{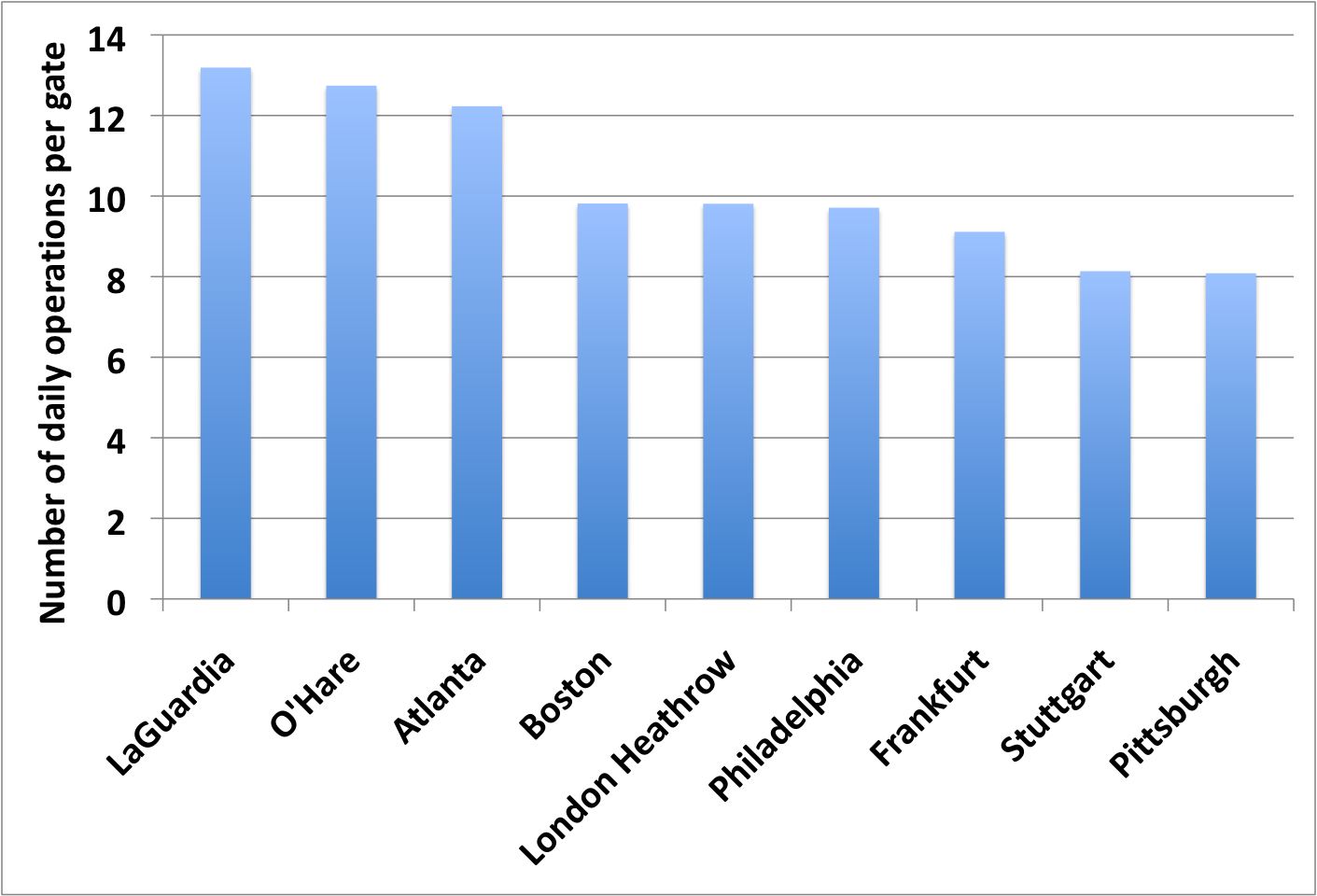}
	\caption{The Number of Daily Operations per Gate for Some Busy International Airports}
	\label{f:dailyOps}
\end{figure}

Two gate assignments are assessed: the current gate assignment and the robust gate assignment. The current gate assignment is obtained from the carrier A, and the robust gate assignment is given in \cite{kim2013atl}. The relationship between taxi-out times, gate-holding times, and $N^*$ is illustrated in Figs.~\ref{f:atl_simTaxiCurrent}-\ref{f:atl_simTaxiRobust}. From the figures, $N^*$ is set to 33. Table~\ref{t:atl_impact} compares the impact of the current gate assignment on gate holding with that of the robust gate assignment. For both gate assignments, it is shown that the reduction of taxi-out times is about 2 min by holding some departures at the gates for 7-8 min. The results in Table~\ref{t:atl_impact} are similar to those in Table~\ref{t:impact}, but the reduction of occurrence of gate conflict is relatively small. Two gate conflicts are eliminated by the robust gate assignment when gate holding is active. The hub airport utilizes two runways for departures and three runways for arrivals and has a large taxiway system as opposed to LGA with one runway shared by departures and arrivals and a small taxiway system. Hence, the hub airport is capable to handle more traffic on the airport surface than LGA, and the saturation of departure throughput occurs at large $N^*$. As a result, about one-fourth of the departures are held at the gates, as compared to more than half the departures held at the gates at LGA. It is concluded that about 75 \% of the departures are cleared to push back before the runways are saturated, and the hub airport is less impacted by gate conflicts than LGA. Fig.~\ref{f:dailyOps} shows the number of daily operations per gate for some busy airports in the world. It is indicated that gates of LGA are busier than those of U.S. major hub airports such as Hartsfield-Jackson Atlanta Airport and Chicago O'Hare Airport. Considering the fact that single runway is used for departures at LGA and multiple runways are used for departures at U.S. major hub airports, the runway throughput of LGA is lower than those of U.S. major hub airports. Lower runway throughput and higher gate utilization at LGA indicate that gate holding would cause more gate conflicts at LGA, so the benefit of the robust gate assignment is relatively small for the hub airport of interest.

\begin{table*}[!t]
	\caption{Impact of Gate Assignment on Gate Holding at a U.S. Hub Airport}
	\label{t:atl_impact}
	\centering
	\begin{tabular}{|c|c|c|c|c|}
		\hline
		& \multicolumn{2}{|c|}{Current Gate Assignment} & \multicolumn{2}{|c|}{Robust Gate Assignment} \\ \hline
		& No Gate Holding & Gate Holding ($N^* = 33$) & No Gate Holding & Gate Holding ($N^* = 33$) \\ \hline
		Number of Gate Conflicts & 12.6 & 33.5 & 3.9 & 31.4 \\
		Number of Gate-held Departures & 0 & 295.2 & 0 & 345.6 \\
		Mean Gate-holding Times & 0 min & 8.4 min & 0 min & 6.9 min \\
		Mean Taxi-out Times & 20 min & 18.3 min & 20.4 min & 18.4 min \\ \hline
	\end{tabular}
\end{table*}

\section{Conclusion}
This paper analyzes the impact of gate assignment on gate-holding departure control. In order to simulate the airport departure process, a queuing model is proposed, consisting of a push-back queue, a taxi-out time estimator, and a runway queue. The model is validated and reproduces airport departure throughput close to the data. Because the performance of gate holding relies on gate separations, a robust gate assignment is introduced.

The results show that gate holding shifts some taxi-out times to gate delays, and it causes gate conflicts between the gate-held departures and arrivals. The robust gate assignment reduces the occurrence of gate conflicts under gate-holding departure control strategies and helps the control strategies to utilize gate-holding times to an extent by maximizing gate separations.

Future work will address the impact of the changed gate assignments on passengers. Most passengers at hub airports are there to transfer their flights as opposed to LGA. Therefore, the change of gate assignments influences passengers' transit time at the airport terminal. Especially, the consequence will be significant to transfer passengers because their transit routes depend entirely on the gate assignments.

\appendices

\bibliographystyle{IEEEtran}
\bibliography{IEEEabrv,iga}

\begin{IEEEbiographynophoto}{Sang Hyun Kim}
is a Ph.D. candidate in the School of Aerospace Engineering at the Georgia Institute of Technology. He holds his B.S. degree from Seoul National University, South Korea. Sang Hyun Kim’s research interests are optimization, transportation, airport operations, ramp management, and general aerospace engineering.
\end{IEEEbiographynophoto}

\begin{IEEEbiographynophoto}{Eric Feron}
received the B.S. degree in applied mathematics from Ecole Polytechnique, Palaiseau, France, the M.S. degree in computer science from the Ecole Normale Sup\'erieure, Paris, France, and the Ph.D. degree in Aeronautics and Astronautics from Stanford University, Stanford, CA. He is the Dutton-Ducoffe Professor of Aerospace Software Engineering, Georgia Institute of Technology, Atlanta. He is also a consulting professor with the Ecole Nationale de l'Aviation Civile, Toulouse, France. Prior to that, he was with the faculty of the Department of Aeronautics and Astronautics, Massachusetts Institute of Technology, Cambridge, for 12 years. His former research students are distributed throughout academia, government, and industry. He has published two books and several research papers. His research interests include using fundamental concepts of control systems, optimization, and computer science to address important problems in aerospace engineering such as aerobatic control of unmanned aerial vehicles, multiagent operations such as air traffic control systems, and aerospace software system certification.
\end{IEEEbiographynophoto}

\end{document}